\input amstex
\documentstyle{amsppt}
\loadbold
\define\bQ{\bold Q}
\define\bq{\bold q}
\define\bv{\bold v}
\define\bJ{\bold J}
\define\bp{\bold p}
\define\bE{\bold E}
\define\bDelta{\boldsymbol \Delta}
\define\bn{\bold n}
\define\bD{\bold D}
\define\bsigma{\boldsymbol \sigma}
\define\bu{\bold u}
\define\bB{\bold B}
\define\bt{\bold \times}

\magnification=\magstep1

\openup2\jot

\topmatter
\title
%%StartAbstract
Steady-State Electrical Conduction in the Periodic Lorentz Gas
%%StopAbstract
\endtitle
\author
%%StartAbstract
N. I. Chernov, G. L. Eyink, J. L. Lebowitz and Ya. G. Sinai
%%StopAbstract
\endauthor
\abstract
%%StartAbstract
We study nonequilibrium steady states in the Lorentz gas of periodic
scatterers when an external field is applied and the particle kinetic energy
is held fixed by a ``thermostat'' constructed according to Gauss' principle  
of least constraint ( a model problem previously studied  
numerically by Moran and Hoover). The resulting dynamics is reversible and 
deterministic, but does not preserve Liouville measure. For a sufficiently 
small field, we prove the following results: (1) existence of a unique
stationary, ergodic measure obtained by forward evolution of initial 
absolutely continuous distributions, for which the Pesin entropy
formula and Young's expression for the fractal dimension are valid; 
(2) exact identity of the steady-state 
thermodyamic entropy production, the asymptotic decay of the Gibbs entropy 
for the time-evolved distribution, and minus the sum of the Lyapunov 
exponents; (3) an explicit expression for the full
nonlinear current response (Kawasaki formula); and (4) validity of 
linear response theory and Ohm's transport law, including the Einstein
relation between conductivity and diffusion matrices. Results (2) and
(4) yield also a direct relation between Lyapunov exponents and zero-field 
transport (=diffusion) coefficients. Although we restrict ourselves here
to dimension $d=2,$ the results carry over to higher dimensions and
to some other physical situations: e.g. with additional external magnetic 
fields. The proofs use a well-developed theory 
of small perturbations of hyperbolic dynamical systems and the method of 
Markov sieves, an approximation of Markov partitions. In our context we
discuss also the van Kampen objection to linear response theory, which,
we point out, overlooks the ``structural stability'' of strongly
hyperbolic flows.
%%StopAbstract
\endabstract
\endtopmatter
\document

\head
Physical Discussion and Statement of Results 
\endhead

\noindent {\it (a) Introduction}

We consider in this paper a dynamical system which corresponds to the 
motion of a single particle between a finite number of fixed,
disjoint, convex scatterers in a periodic domain of the plane 
${\Bbb R}^2.$ As in the previous works [3,4], the particle changes 
its velocity at moments of collision according to the usual law of 
elastic reflection, but, unlike there, the particle motion between 
collisions is not the free one at constant velocity. Instead, 
the motion between collisions is governed by the following set of 
first-order equations:
$$
\dot{\bq}=\bp/m \tag 1
$$
$$
\dot{\bp}=\bE-\zeta\cdot \bp \tag 2
$$
Here, $\bq=(q_1,q_2)$ are the Cartesian coordinates of the particle,
and $\bp=(p_1,p_2)$ the corresponding momenta. $E$ is a constant
electric field and the ``friction coefficient'' $\zeta$ is chosen 
as a phase space function so that the kinetic energy (or speed) of 
the particle is conserved:
$$
\zeta=  \bE\cdot(\bp/m)/(p^2/m). \tag 3
$$
We shall regard the kinetic energy of the particle as defining a 
``temperature'' according to the relation $p^2/m=k_B T=1/\beta$. 
Because of the conservation of kinetic energy, we may consider 
the reduced phase space at each value of the particle speed $v$, 
with coordinates $X=(q_1,q_2,\theta),$ where $\theta$ is the angle 
of the particle velocity vector with respect to the 1-direction. 
It is an elementary calculation that 
$$
\nabla_{X}\cdot \dot{X}=-\zeta, \tag 4
$$
so that the Liouville measure is not preserved when $E\neq 0.$
On the other hand, observe that Eqs.(1-2) define a flow $\{ S^t_E\}$ on the 
phase-space, running backward as well as forward, and that the 
inversion $\tilde{X}=(q_1,q_2,\theta-\pi),$ corresponding
to velocity-reversal, has the property that $S^t_E(\tilde{X})
=(S^{-t}_EX)\tilde{\,}.$

The model under consideration was previously studied theoretically
and numerically in [30]. It is a simple example for a set of new methods
in non-equilibrium molecular dynamics (NEMD) which has been developed
in the past decade by W. G. Hoover, D. J. Evans, G. P. Morriss and
others [13,21]. Unlike a more traditional approach (see, for
example, [28]) which models interactions of the physical system
with a heat bath by including suitable stochastic elements in the 
dynamics, the new techniques are based upon dynamics which are purely 
deterministic and reversible but for which the Liouville theorem is invalid.
In one version of the method, which we study, the total
kinetic energy of a system of particles subjected to a thermodynamic 
or mechanical driving field is held fixed by modifying the dynamics 
according to a prescription of Gauss, the ``principle of least
constraint'' [17]. What Gauss proposed was that forces $\bold F^{(c)}_i$
be added to the Newtonian dynamics in such a way that a chosen constraint
$f(\bq,\dot{\bq},t)=0$ be maintained and the total magnitude
of the (Jacobi frame) constraint force,
$$
 \sum_{i=1}^N \,\, (F^{(c)}_i)^2/m_i,
$$
be minimized instantaneously. In the space of forces (or accelerations)
the constraint defines a hyperplane by a linear equation $ \sum_{i=1}^N 
{ \bn}_i({ \bq},{\dot{ \bq}},t)\cdot{ a}_i=b({ \bq},{\dot{ \bq}},t),$
with  ${ \bn}_i({ \bq},{\dot{ \bq}},t)= \nabla_{\dot{{ \bq}_i}}f({ \bq},
{\dot{ \bq}},t).$ Since ${\bold F}_i^{(c)}\propto { \bn}_i$ by Gauss'
principle, the final equations of motion are of the form:
$$
m_i\ddot{{ \bq}}_i={\bold F}_i -\zeta { \bn}_i,
$$
for some $\zeta.$ Obviously, our dynamics is a special case of this 
general construction. For a system of many particles, holding fixed the 
``peculiar'' kinetic energy, $K={{1}\over{2}}\sum_{i=1}^N m_i({\dot{
\bq}}_i-\bu_i)^2$ ($\bu_i$ is the expected velocity of particle $i$), 
should be equivalent to holding the temperature fixed, according 
to the identification $K=N\cdot d\cdot k_B T/2.$ In our simple example with
$N=1$ the identification of ``temperature'' is not really appropriate, but we
use the term by analogy. 

In practical simulations with realistic potentials and also in the 
simulations for our simple model in [30], the empirical measure
$$
\mu^{X}_t\equiv {{1}\over{t}}\int_0^t ds\,\,\delta_{S^sX}
$$
appears to converge (weakly) to a final stationary distribution for
almost every initial point $X$ of the phase space. However,
the mathematical proof of existence and uniqueness of stationary
measures, which can be given for some cases of the traditional
stochastic approach [18,19], is generally lacking here. Furthermore,
in contrast to the stochastic modeling method, the stationary
distributions appear on the basis of numerical evidence to be
singular with respect to Liouville measure. Indeed, the measures
appear to be { multifractal}, with an information dimension 
strictly less than the dimension of the constraint surface in phase 
space([13], Ch.10). The simple model we consider affords the opportunity 
to rigorously examine such issues.

In addition, simple formal arguments suggest some remarkable 
properties of the Gaussian dynamics. It is found in particular that
the physical entropy production in the steady state is just equal
to the asymptotic rate of decrease of the Gibbs entropy for 
time-evolved initial distributions, and the latter is seen
to be just the negative of the sum of the Lyapunov exponents (defined
almost surely with respect to the final stationary measure) [13]. 
>From this results immediately a relation between the transport coefficients,
which appear in the entropy production, and the Lyapunov exponents
for the Gaussian dynamics. We wish also to study the validity of such 
relations in our simple model example.

The plan of the paper is as follows: In the next section of Part I we 
give simple formal arguments -- which are later made into proofs
-- for the relations mentioned above, as well as indicate some
generalizations and extensions which we do not prove afterward in 
all details. In the section following that, we
give precise formulations of the rigorous results we establish for the
model, and indicate some basic ideas of the argument. In addition,
we point out the surprising transformation of this essentially 
non-equilibrium problem into a problem of (lattice) equilibrium
Gibbs measures, by the method of Markov partitions and symbolic
dynamics. In the final section of Part I we analyze in detail the
failure in our model of the van Kampen argument against validity 
of linear response theory. In the more technical Part II (which is
due essentially to N. C. and Y. S.) the proofs of all the main results
are outlined.

\noindent {\it (b) Formal Arguments}

Let us first give the argument for the relations between physical entropy
production, time-derivative of Gibbs entropy, and Lyapunov exponents.
Although our discussion is entirely in the context of the Lorentz
model, it will appear that the basis of the results is rather general
[13]. We assume in our discussion that, if $\mu$ is an initial measure
absolutely continuous with respect to the Lebesgue measure and $\hat{S}^t_E\mu=
\mu\circ S^{-t}_E,$ then $\hat{S}^t_E\mu\rightarrow\mu^{+}_E$ (weakly) as
$t\rightarrow +\infty,$ where $\mu^{+}_E$ is the physical stationary
measure for the forward evolution. For simplicity we denote $\hat{S}^t_E\mu$
by $\mu_t$ and its density with respect to Liouville measure by $f_t.$
With the usual definition of Gibbs entropy,
$$
S(\mu)= -k_B\int f(X)\log f(X)\,\, dX,
$$
it is a simple calculation for differentiable $f$ that
$$
{{d}\over{dt}}S(\mu_t)=k_B\int\,\,f_t(X)
 (\nabla_X\cdot \dot{X})\,\,dX
 =-k_B\mu_t(\zeta),
$$
using in the last equality Eq.(4). Observing from Eq.(3) that 
$\zeta$ is a bounded, continuous function on the phase space,
we can therefore infer that
$$
\lim_{t\rightarrow +\infty}
-{{d}\over{dt}}S(\mu_t)= k_B\mu^{+}_E(\zeta), 
$$
and, in fact, the right side is just equal to $\mu^{+}_E({ \bv})\cdot
{ \bE}/ T.$ This has an interesting physical interpretation: if we
consider ${ \bJ}\equiv\mu^{+}_E({ \bv})$ as the steady state
electrical current, then it is just ${ \bJ\cdot \bE}/ T,$ which is 
the entropy production due to Ohmic dissipation [20]. (It may
appear odd that the {\it negative} time-derivative of Gibbs entropy
corresponds to entropy production. One should understand that 
the Gaussian dynamics are supposed to model the effect of reservoir
elements on the particle system, for which the { total} system,
reservoir+particles, obeys the Liouville theorem. Hence, the 
decrease of particle entropy corresponds to the increase of 
reservoir entropy, and the latter represents the physical entropy 
production.)   

For the other half of the relation, we note that, if there exist 
local stable and unstable subspaces in the tangent space to ${\frak M}$
(for every point where the flow is smooth), then one may define local 
exponential rates of contraction, $\Lambda_E^s(X),$ and expansion, 
$\Lambda_E^u(X),$ along those one-dimensional subspaces. (The third 
direction along the flow is neutral.) The volume of a small
parallelopiped with one leg along the flow direction and the other
legs along the stable and unstable rays has the volume which is the
triple wedge product of those legs, i.e. the product of their
magnitudes and a combination of trigonometric functions of the angles 
between the legs. We denote the latter angular factor by $U_E(X),$
defined for all points of smoothness of the flow. Then, there holds
the following relation for all $X$
$$
(\nabla_{X}\cdot \dot{X})(X)=\Lambda_E^s(X)+
    \Lambda_E^u(X)+ \frac{d}{dt}U_E(S_E^tX)|_{t=0},\,\,\,\,\,\,\,\,\,
$$
except for the singular points of the flow.
On the other hand, $\lambda^s_E=\mu^{+}_E(\Lambda_E^s)$ and 
$\lambda^u_E=\mu^{+}_E(\Lambda_E^u)$ are the Lyapunov exponents
for the ergodic measure $\mu^{+}_E, $ while the time-derivative
has zero expectation just by stationarity under $S_E^t.$
Hence, we have also the relation
$$
{{1}\over{T}} { \bE\cdot \bJ}= -k_B(\lambda^s_E+\lambda^u_E). \tag 5
$$
The content of this remarkable relation is the equality of 
the full (nonlinear) entropy production and the negative sum of the
Lyapunov exponents. 

We now turn to a discussion of the formal response theory. It is 
very useful here to develop certain exact integral expressions for the 
stationary measures $\mu^{+}_E.$ To this end, let us note that 
the density of the measure at time $t$ starting from initial Lebesgue
measure is just given by the Jacobian determinant
$$
 f_t(x)=\left| {{\partial S^{-t}_E
(X)}\over{\partial X}}\right|.
$$
Then, by an elementary calculation,
$$
\partial_t f_t(X)=-(\nabla_{X}\cdot \dot{X})|
_{S^{-t}_E X}\cdot f_t(X)= \beta({ \bE\cdot \bv}(S^{-t}_E X))
\cdot f_t(X), \tag 6
$$
and, by direct integration,
$$
f_t(X)=1+\beta { \bE}\cdot\int_0^t\,\,{ \bv}(S_E^{-s}X)
f_s(X)\,\,ds. 
$$
Therefore, for any well-behaved function $\phi$ on the phase space, 
$$
\hat{S}_E^t\mu_0(\phi)=\mu_0(\phi)+\beta{ \bE\cdot}\int_0^t\mu_0
({ \bv}\cdot(\phi\circ S_E^s))\,\,ds 
$$
(where, note, $\mu_0$ according to our previous notation is just 
Lebesgue measure, which is stationary for $E=0.$) By our
assumption, $\hat{S}_E^t\mu_0$ converges to $\mu_E^{+}$ as
$t\rightarrow +\infty,$ and, if the integrand on the right side has
sufficiently good decay, we obtain finally
$$
\mu_E^{+}(\phi)=\mu_0(\phi)+\beta{ \bE}\cdot\int_0^{\infty}
\mu_0({ \bv}\cdot(\phi\circ S_E^t))\,\,dt. \tag 7
$$
Such exact expressions for steady-state measures are sometimes called 
in the physics literature ``non-equilibrium statistical distributions'' 
and have been known for a long time [27,29,44]. Assuming that ${\bv}$
is in the class of $\phi$ for which the expression is valid, one
obtains at once a formula for the exact current-response as a nonlinear
function of field:
$$
{ \bJ}({ \bE})=\beta{ \bE\cdot}\int_0^{\infty}\mu_0({ \bv}\otimes
({ \bv}\circ S_E^t))\,\,dt. \tag 8
$$
This is an example of the so-called {\it Kawasaki formula} for the
nonlinear response [41]. As we see below, it may be regarded as
a generalization of the usual Green-Kubo formula.

It is now easy to specialize the above results to obtain the response
to linear order in the field. Indeed, we see formally that the linear 
correction term to the measure $\mu_E^{+}$ is just given by
$$
\mu_E^{+}(\phi)=\mu_0(\phi)+\beta{ \bE \cdot}\int_0^{\infty}
\mu_0({ \bv}\cdot(\phi\circ S_0^t))\,\,dt+o(E). \tag 9
$$
This is rigorously correct, for example, if there is a bound on the
decay in $t$ of $\mu_0(\bv\cdot(\phi\circ S_E^t))$ uniform in $E,$ so that 
dominated convergence may be applied to show the remainder term is 
really $o(E).$ Needless to say, this is not just a fine point of
rigor but is exactly where dynamical properties enter in the
derivation of the transport law. In fact, from Eq.(9) one 
obtains directly {\it Ohm's law}
$$
{ \bJ}={ \bsigma}\cdot{ \bE}+o(E) \tag 10
$$
with 
$$
\bsigma=\beta\int_0^{\infty}\mu_0((\bv\circ S_0^t)\otimes\bv)\,\,dt.
$$
The latter yields immediately the {\it Einstein relation}
$$
\bsigma=\beta\cdot \bD\tag 11
$$
where ${\bold D}$ is given by the usual Green-Kubo formula.
Furthermore, putting together Eqs.(5),(10), and
(11), and assuming for the moment isotropy, we obtain a 
simple formula for the diffusion coefficient as:
$$
D=\lim_{E\rightarrow 0}{{-k_B T^2(\lambda^s_E+\lambda^u_E)}\over
{E^2}}. 
$$
The relation appears naturally as both transport coefficient and
Lyapunov exponents are related to entropy production. 

There is nothing in the above arguments which imposes a restriction
to $d=2,$ and corresponding results for higher dimensions can be
rigorously obtained by using extensions of our methods developed
in [10]. Another interesting generalization is to consider the 
addition of an external {\it magnetic field} to the dynamics.
This involves just the addition of an appropriate Lorentz force
to the left side of Eq.(2):
$$
\dot{\bp}=\bE+\bp\bt\bB-\zeta\cdot \bp.
$$
Because the magnetic interaction is Hamiltonian and conserves 
kinetic energy, the definition of $\zeta$ is the same and also
the formula (4) for the divergence of the dynamical vector-field
remains valid. It is therefore easy to derive formally in the same 
manner as before, expressions like Eqs.(7) and (8), with only
$S^t_E$ replaced by $S^t_{B,E}.$ In particular, the Kawasaki-type
formula for current response remains valid:
$$
{ \bJ}(\bB,\bE)=\beta{ \bE\cdot}\int_0^{\infty}\mu_0({ \bv}\otimes
({ \bv}\circ S_{B,E}^t))\,\,dt.
$$
The above formula decribes several new phenomena that arise
in the simultaneous presence of electric and magnetic fields,
e.g. the {\it Hall effect} of transverse electrical currents. 
>From this follows also validity of Ohm's law, as in Eq.(10),
but with a $B$-dependent conductivity given by 
$$
\bsigma(\bB)=\beta\int_0^{\infty}\mu_0((\bv\circ S_{B,0}^t)\otimes\bv)\,\,dt.
\tag 12
$$
Although time-reversal symmetry is broken by the magnetic field,
one can derive from this expression by considering time-inversion
the relation
$$
\bsigma^{\top}(-\bB)=\bsigma(\bB),
$$
which is the usual {\it Onsager-Casimir reciprocal
relation} for the transport coefficient.

\noindent {\it (c) Rigorous Results}

We now state precisely the results along the above lines that we can 
establish by---we emphasize---essentially just supplying necessary rigor to 
the above formal arguments. The basic properties we need, as we have 
seen, are existence of local stable and unstable manifolds,(weak) 
convergence to a stationary, ergodic measure and some uniform
decay of correlations. The general approach to rigorous derivation of 
the transport law is the same as that given in [14] for the case of
stochastic lattice gases. Here the necessary information on uniform 
correlation bounds is established by approximating the deterministic
dynamics by a suitable Markov chain, through the method of {\it Markov
sieves.} Unfortunately, the formal argument as we have given it above
cannot be presently made rigorous, since there are so far no proofs
of decay of correlations for the Lorentz gas (even at zero field)
in ``true'' time. Instead, such bounds have been obtained by counting
time in terms of numbers of successive collisions. This is connected
with the so-called {\it special representation} of the flow, a
particular application of a general technique of ergodic theory.
We give now a more precise formulation of our model than in the 
Introduction (partly to set notations) and then describe the special
representation. Afterward, we formulate our main results as a series 
of propositions with brief indications of the main ideas of the 
proofs and, finally, discuss the notions of Markov partitions
and symbolic dynamics which are not actually used in the proofs but
give some additional insight into the model.

As stated in the Introduction, we consider the system
of a moving particle in the torus ${\Bbb T}^2$ with a finite number
of disjoint strictly convex scatterers. The region of that torus
complement to the union of all the scatterers is denoted
by $Q$. Recall that the velocity of the moving particle
is constant and equals $v=p/m$. Therefore, the phase
space of the system is now $\frak M = Q\times S^1_v$ where
$S^1_v$ stands for the circle of radius $v$. The motion
under the field $E$ generates the phase flow $\{ S^t_E\}$
in $\frak M$ where $-\infty <t< \infty$. Define a subset of $\frak M$
corresponding to the points of collision of the particle with
the boundary $\partial Q$:
$$
  M = \{ x=({ \bq},{ \bv})\in\frak M: { \bq}\in\partial Q, { \bv}
 \in S^1_v, { \bv}\cdot{\bold n}({ \bq})\geq 0\}
$$
Here and further on ${\bold n}({ \bq})$ stands for the inward unit normal
vector to $\partial Q$ at the point ${ q}$; we have chosen rather 
arbitrarily to label points by their velocity the instant after 
collision.  Standard coordinates in $M$ are $r$ and $\varphi$
[5,6], where $r$ is the arc length parameter along $\partial Q$ 
and $\varphi$ is the angle between ${ n}({ q})$ and ${ v}$ at a point 
$x=({ q},{ v})\in M,\,\, |\varphi|\leq\pi /2$. Rather
than $\varphi$ it is sometimes useful to consider the coordinate
$s=\sin\varphi,\,\,|s|\leq 1.$ A map $T_E: M\rightarrow M$ may be defined
by taking each point $x\in M$ to the point of its next collision, the
so-called {\it billiard ball map} or 
{\it first-return map.} We denote the time until the next 
collision by $\tau_E(x)$ and note the important restriction of all
our considerations to the case of {\it finite horizon}. In that case 
$\tau_E(x)\leq \tau_{\max}<+\infty$ except possibly on the codimension 1 
singularity set  $S_{-1}=T_E^{-1}(\partial M)$ of the piecewise smooth
map $T_E$ (where $\partial M=\{x=({ \bq},{ \bv})\in M: { \bv}\cdot{\bold n}
({\bq})=0\}$.) However, the singularities of $T$ are mild -- they cannot 
accumulate, i.e. after a finite number of collisions the trajectory
will escape the vicinity of $S_{-1}$ and stay uniformly far from it.
The map $T_E$ has an inverse $T_E^{-1}$ which is 
also piecewise smooth with a singularity set $S_1=T_E(\partial M).$
On the space $M$ there is defined the inversion $\tilde{x}=(r,-\phi)$
for $x=(r,\phi)$ under which $T_E(\tilde{x})=(T_E^{-1}x)\tilde{\,}.$
Note that the billiard flow
$\{S^t_0\}$ preserves the Liouville measure $d\mu_0=d{ \bq}\,d{ \bv}$
and the billiard map $T_0$ preserves the measure
$d\nu_0=dr\,ds=\cos\varphi dr\,d\varphi$.

Since the dynamics for any E is completely
deterministic, it is obvious that any point $({ \bq},{ \bv})\in \frak M$
is completely specified by giving the point $x\in M$ corresponding to
its state just before its last collision and the time $\tau$ since
that last collision. This is the so-called special representation of
the flow. More formally, it corresponds to representing the system 
$(\frak M, S_E^t)$ as the flow under the (ceiling) function $\tau_E$ 
generated by the map $T_E$ on the (base) space $M;$ e.g. see the 
general discussion in Chapter 11 of [11]. Let us just remark here
that if $\nu_E$ is a measure on $M$ invariant (ergodic) under $T_E,$ then 
a measure $\mu_E$ on $\frak M$ invariant (ergodic) under $S_E^t$ is
defined by 
$$
\mu_E(f)=\frac{1}{\bar{\tau}_E}{\int_M\,\nu_E(dx)\int_0^{\tau_E(x)}\,d\tau\,\,
 f(x,\tau)}, \tag{13}
$$
for $f\in C(\frak M),$ using the isomorphism $\frak M\cong
\{(x,\tau): x\in M,0\leq \tau\leq \tau_E(x)\}.$ 
Here, $\bar{\tau}_E\equiv\nu_E^{+}(\tau_E),$ the mean collision time,
appears as a denominator for proper normalization.  
Notice that the measures $\mu_0,\nu_0$ previously defined are 
in fact so related. There is also a simple relation between the 
Lyapunov exponents $\tilde{\lambda}^{u,s}_E$ for any ergodic 
measure $\nu_E$ under $T_E$ and those for the associated $\mu_E,$ which 
is just $\lambda^{u,s}_E=\tilde{\lambda}^{u,s}_E/\bar{\tau}_E.$

We can now state our results. They all require the condition of finite
horizon and sufficient smallness of the field, which we hereafter
assume without explicit mention. First, we have (defining, as above,
$\hat{T}^n_E\nu\equiv\nu\circ T_E^{-n}$)

\proclaim{Proposition 1} There is a stationary, ergodic measure
$\nu^+_E$ for the map $T_E$ on $M$, which is the weak limit
$$
\nu_E^+=w-\lim_{n\rightarrow +\infty}\hat{T}^n_E\nu,
$$
for any measure $\nu\ll\nu_0$ on $M.$ The measure $\nu_E^+$ 
satisfies the Pesin entropy formula
$$
h_{\nu_E^+}(T_E)=\tilde{\lambda}_E^u,
$$
and Young's formula for the fractal (or information) dimension
$$
HD(\nu_E^+)= h_{\nu_E^+}(T_E)\left(\frac{1}{\tilde{\lambda}^u_E}-
             \frac{1}{\tilde{\lambda}^s_E}\right).
$$
Furthermore, considering $S_E^t$ as the
special flow, one obtains by the construction in Eq.(13)
above, an invariant ergodic measure $\mu_E^{+}$ which is the weak limit
$\mu_E^{+}=w-\lim_{t\rightarrow +\infty}S_E^t\mu$ for any
$\mu\ll\mu_0,$ which 
has the entropy given by $h_{\mu_E^+}(S_E^1)=h_{\nu_E^+}(T_E)/\bar{\tau}_E$
and the fractal dimension $HD(\mu_E^+)=HD(\nu_E^+)+1.$
\endproclaim

This result is proved in Part II, Sections (d) and (f). It is easy to 
check by combining the Pesin and Young formulas that
$$
HD(\mu_E^+)=2+\left(1+\frac{\bar{\zeta}_E}{h_E}\right)^{-1}, \tag 14
$$
where $h_E\equiv h_{\mu_E^+}(S^1_E)$ and 
$\bar{\zeta}_E\equiv -(\lambda^u_E+\lambda^s_E).$ The next main result
allows us to physically identify the latter expression:

\proclaim{Proposition 2} For any measure $\mu$ on $\frak M$, 
$\mu\ll\mu_0,$ 
$$
{ \bJ}\cdot{ \bE}/T=-\lim_{t\rightarrow +\infty}{{d}\over{dt}}
    S(\mu_t)=-k_B(\lambda^u_E+\lambda^s_E).
$$
Here, as above, ${ \bJ}=\mu_E^{+}({\bv}),\mu_t=\hat{S}_E^t\mu$
and $\lambda^u_E,\lambda^s_E$ are Lyapunov exponents for the
ergodic measure $\mu_E^+.$
\endproclaim

This result follows from the weak convergence in Proposition 1
and the existence of stable and unstable subspaces in ${\frak T}_X{\frak M}$ 
except for the codimension 1 set of points $X\in {\frak M}$ where
the flow is singular. The existence of the subspaces is guaranteed by 
the strong hyperbolicity of the flow $S_E^t,$ which holds for 
small $E$ (see Part II and Ref.[24]). It is perhaps
interesting to note that the ratio determining the fractal dimension
is just the physical entropy production $\bJ\cdot\bE/T$ divided by 
($k_B$ times) the dynamical entropy production $h_E$ 

To state the next results, we must introduce some additional notation.
For any $X\in{\frak M}$ let us define ${ \bQ}(X)$ to be the
projection onto the configuration space $Q,$ i.e. ${ \bQ}({ \bq},
{ \bv})={ \bq}.$ In particular, ${ \bQ}(x)\in \partial Q$ for 
$\in M.$ We then define on $M$ a function ${ \bDelta}_E$ by
$$
{ \bDelta}_E\equiv { \bQ}\circ T_E-{ \bQ}.
$$
Clearly, ${ \bDelta}_E(x)$ is just the total vector displacement of the
particle from its starting point $x\in M$ until its next collision.
This definition is actually ambiguous since we use periodic 
boundary conditions on $Q$. For the results below it is crucial 
that $\bDelta_E$ be defined by the 
convention that, when a particle in ${\Bbb R}^2$ crosses the boundary
of the fundamental domain $Q$, the displacement is evaluated between the
starting point in $Q$ and the final position in the adjacent domain
(rather than its periodic image in the fundamental domain.)
Let us also define a space of H\"{o}lder continuous functions
$H_{\alpha}=\{f:|f(x)-f(y)|\leq C_f|x-y|^{\alpha}\,\,
\text{for any }x,y\in M\}.$
More generally, let $H_{\alpha}^{*}$ denote the space of {\it piecewise}
H\"{o}lder continuous functions, which are H\"{o}lder continuous 
(with an exponent $\alpha$) on a finite collection of subdomains in $M$
separated by a finite union of compact smooth curves. (Note that the
curves and domains must be fixed for the class $H_{\alpha}^{*}$ under
consideration: we consider below the case where the discontinuities
occur on the singularity sets $S_{-1,1}=S_{-1}\cup S_1$ of both the maps 
$T_E^{\pm}$ and $T_0^{\pm}$.) Then, an exact integral formula for invariant
expectations and a Kawasaki-type formula for nonlinear response can 
be established in the following form:

\proclaim{Proposition 3} For any $\phi\in H_{\alpha}^{*}$,
$$
\nu_E^{+}(\phi)=\nu_0(\phi)+\sum_{n=1}^{\infty}\nu_0
     [(\phi\circ T_E^{n})(1-e^{-\beta { \bE\cdot\bDelta}_E})].
    \tag 15
$$
Furthermore, the following equality holds
$$
{ \bJ}({ \bE})={{1}\over{2\bar{\tau}_E}}
    \nu_0[{ \bDelta}_E(1-e^{-\beta { \bE\cdot\bDelta}_E})]  
    +{{1}\over{\bar{\tau}_E}}\sum_{n=1}^{\infty}              
 \nu_0[({ \bDelta}_E\circ T_E^n)(1-e^{-\beta { \bE\cdot\bDelta}_E})]. 
$$ 
\endproclaim

The proof of this Proposition is obtained by repeating the formal
arguments given earlier, but now for the map $T_E$ (see Part II (a).)
To make the argument rigorous it must be shown, for the first
Eq.(15), that the function $f_E(x)=1-e^{-\beta { \bE\cdot\bDelta}_E (x)}$
obeys $\nu_0(f_E)=0$ and is in $H_{\alpha}^{*}$ 
for all (small) E: the result then follows from 
Theorem 17 of Part II(e) which proves the decay of correlations to 
guarantee convergence of the summation. The second Eq.(16) follows 
more or less directly from the first one (see II(a)).

Finally, we have the following results on the linear response,
transport law and Einstein relation:

\proclaim{Proposition 4}  For any $\phi\in H_{\alpha}^{*}$,
$$
\nu_E^{+}(\phi)=\nu_0(\phi)+\beta{ \bE\cdot}\sum_{n=1}^{\infty}
    \nu_0[(\phi\circ T_0^{n}){ \bDelta}_0]+o(E). \tag 16
$$
Furthermore, ${ \bJ}=\bsigma\cdot{ \bE}+o(E)$ where $\bsigma=\beta{ \bD}$
and
$$
{ \bD}=\frac{1}{2\bar{\tau}_0} \sum_{n=-\infty}^{+\infty}
    \nu_0(({ \bDelta}_0\circ T_0^n)\otimes{ \bDelta}_0). \tag 17
$$
\endproclaim

The latter expression is just the (discrete-time) Green-Kubo formula
established by Sinai and Bunimovich in [4] for the diffusion
coefficient. See also [6], where this formula was corrected
and the matrix $\bD$ was shown to be positive-definite.
In these works, the diffusion coefficient is naturally 
defined through the covariance of the limiting Wiener
process for the rescaled particle motion in the Lorentz array of scatterers.
Thus, the relation $\bsigma=\beta \bD$ is a proper form of the
Einstein relation. (Although we give no details here, similar results
can be obtained for the situations with an external magnetic field,
as long as the scatterer array has finite-horizon for the motion
along circular arcs produced by the Lorentz force. The limiting
Brownian motion with $\bB$-dependent drift can be obtained, for  
example, by the arguments in [10], and the diffusion coefficient
is then given just by the above Green-Kubo formula with $T^n_0$
replaced by $T^n_{B,0}.$ In that case, therefore, there is a generalized
Einstein relation of the form $\tilde{\bsigma}(\bB)=\beta \bD(\bB),$
where $\tilde{\bsigma}$ is the symmetric part of $\bsigma$
given by the formula (12) or by a similar discrete-time formula.)
The proof of Eq.(16) follows from the {\it uniform}
correlation bounds in PartII(e), which allows dominated convergence to
be applied to show the remainder term is $o(E).$ The rest of the 
Proposition then follows rather directly (see II(a).) From Ohm's law
and the Eq.(14), we can obtain an estimation for the fractal dimension
valid for small $E$:

\proclaim{Corollary}
$$
HD(\mu_E^+)=3-\frac{\bE^{\top}\cdot\bsigma\cdot\bE /T}{k_B\cdot h_0} +o(E^2).
$$
\endproclaim

This requires the proof that $\lim_{E\rightarrow 0}h_E=h_0$
(see Part II (f).) In particular, it follows that $HD(\mu_E^+)<3$ when 
$E$ is small but finite. 

The method of Markov sieves is the main technical tool used both in
the construction of the invariant measure and in the proof of the
correlation bounds. The construction of the measure proceeds by 
first defining in a usual manner a conditional measure $p^c$ on unstable
fibers $\gamma^u(x)$ and, then, for a selected fibre $\gamma^u$, proving both  
the existence of the limit $w-\lim_{n\rightarrow =\infty}\hat{T}_E^np^c$
and its independence of the choice of $\gamma^u.$ Without entering into
any technical details, we just remark that $\hat{T}_E^n p^c(B)$--- for a 
fixed ``parallelogram'' $B$ circumscribed by stable and unstable fibres---
is shown to have a limit only by studying the simultaneous evolution
of the whole collection of parallelograms composing the Markov sieve.  
Likewise, the correlation bounds are obtained by using the Markov
sieve to approximate the entire evolution sufficiently well by a 
Markov chain with good decay of correlations. The uniformity in $E$
is essentially the consequence of ``structural stability'' of the
strictly hyperbolic map $T_0$ to small perturbations: see Section (d)
below.
 
The related method of Markov partitions, which played the crucial role in 
earlier work [3,4], is not used here. Nevertheless, it should be possible 
to construct such partitions with good enough properties, and, not only 
would this yield some further conjectured results, but also some additional
heuristic insights into the model. We therefore very briefly explain 
this method and the related idea of symbolic dynamics. A {\it Markov 
partition} is a certain countable partition $\eta$ of $M$ 
into parallelograms $\eta=\{A_i: i\in  I\} (for the definition of 
parallelograms see our section II, c).$ As a notation, we write 
$\gamma^u_{\eta}(x)=\gamma^u(x)\cap A_i,$ where $x\in A_i\in\eta$ and  
$\gamma^u(x)$ is the local unstable fibre through $x$ (with a similar
notation for the stable fibres.) Then, the Markov partition has the 
fundamental property that for a.e. $x\in M,$
$$
T(\gamma^s_{\eta}(x))\subseteq \gamma^s_{\eta}(Tx),\,\,\,\,
    T^{-1}(\gamma^u_{\eta}(x))\subseteq \gamma^u_{\eta}(T^{-1}x).
$$
It can be shown that the sets of the form $\cap_{n=-\infty}^{+\infty}
T^nA_{\omega_n}$ for any sequence
$\omega=\{...,\omega_{-n},...,$\linebreak $\omega_0,...,\omega_n,...\}\in{\frak
I}^{\Bbb Z}$ may consist only of one or no points, and 
$\cap_{n=-\infty}^{+\infty}T^nA_{\omega_n}\neq\emptyset$ if
$\nu(A_{\omega_n}\cap TA_{\omega_{n+1}})>0$ for all $n\in {\Bbb Z}.$
Setting $\pi_{ij}=1$ if $\nu(A_i\cap TA_j)>0$ and 0 otherwise, we may
define the space $\Omega$ of symbolic sequences $\omega$ as the subset of
${\frak I}^{\Bbb Z}$ such that $\pi_{\omega_n\omega_{n+1}}=1$ for
every $n\in {\Bbb Z}.$ Then it may be seen that the mapping
$\Phi :\Omega\rightarrow M$ defined as
$$
\Phi(\omega)=\cap_{n=-\infty}^{+\infty}T^nA_{\omega_n}
$$
is one-to-one onto a set of full $\nu$-measure. In a natural way it
gives an isomorphism of measurable spaces for $M$ and $\Omega.$
Furthermore, if $S$ is the {\it shift} on $\Omega$ defined by
$(S\omega)_n=\omega_{n-1},$ then $\Phi\circ S=T\circ\Phi$ so that
$\nu$ is pulled back to an $S$-invariant measure $\rho$ on $\Omega.$ 
In fact, the essential point is that $\rho$ is a kind of
one-dimensional lattice Gibbs measure for the spin-system with
values of spin in the countable set ${\frak I}$ and with interactions of
sufficiently rapid spatial decay. The formal Hamiltonian of the
measure $\rho^{+}_E$ (corresponding to $\nu_E^{+}$) is just given by
the expression
$$
H(\omega)\approx\sum_{n=-\infty}^{+\infty}
{\tilde{\Lambda}}^u_E(\Phi(S^n\omega)), \tag 18
$$
in which quantity $\tilde{\Lambda}^u_E(x)$ is the local exponential 
rate of expansion for $x\in M$ under the map $T_E,$ and is simply related
to the expansion rate for $S_E^t$ by $\tilde{\Lambda}^u_E(x)
=\int_0^{\tau_E(x)}d\tau\,\,\Lambda^u_E(x,\tau).$
What is very remarkable about this isomorphism in our context is that
it converts the essentially nonequilibrium measure $\nu_E^{+}$ 
into a lattice Gibbs measure $\rho_E^{+}$ of an equilibrium spin-system!
This  is the main feature of the so-called thermodynamical
formalism (see e.g. [33]).

This transformation, if it can be rigorously carried out here, should
have various consequences. First, it is believed on the basis of 
various numerical evidence that the measure $\nu^{+}_E$ should 
be {\it multifractal}, with an entire continuous spectrum of associated
dimensions. In fact, we expect on general grounds that this 
should be so (for small $E$) and the verification by the above 
method of symbolic representation should be possible along the lines
in [39]. Another interesting consequence of the representation
as a lattice Gibbs measure is that there should be a {\it variational
principle} which characterizes the measure $\nu_E^{+}.$ Specifically,
the measure $\nu_E^{+}$ should minimize the quantity $f(\nu)\equiv\nu(\tilde{
\Lambda}_E^u)-h_{\nu}(T_E)$ among all the ergodic measures for $T_E$ on $M.$
The question of existence of a variational characterization of steady 
state measures has been a traditional one in non-equilibrium
statistical mechanics, with a {\it principle of minimum entropy
production} most often proposed [14]. It may be of some interest 
to observe that linear corrections to invariant measures, such as
in Eq.(16) above, are known to be correctly prescribed by minimizing
entropy production in some simple stochastic dynamical contexts
[14,25]. However, no such principle is known to be exactly valid, and the 
present example is the only one we know where a microscopic measure 
is precisely characterized. It may be that this example is, in fact, a little 
more reminiscent of the {\it maximum entropy principle} proposed by 
Zubarev [45]. On the other hand, it must be admitted
that the form of the principle considered here is quite different from  
any of the ones conventionally considered in non-equilibrium statistical 
mechanics. The minimization is only within the class of ergodic
measures and, while $f(\nu)\geq 0$ in that class, the minimum should
be exactly zero (so that $\nu_E^+$ satisfies the Pesin formula.)
Whereas standard variational principles seek to characterize a (unique)
invariant measure out of the class of all probability measures, here the
principle characterizes a certain {\it physical measure} out of the 
infinite class of ergodic measures for the deterministic dynamics.

A final remark which we make regarding the Gibbsian formulation
of our problem, is that, in this guise, the conductivity $\bsigma$
appears as a ``susceptibility,'' or the equilibrium response to
an applied field. That is, the current ${ \bJ}$ is just the 
expected value of a variable ${\bold A}_E$ (actually, 
${\bold A}_E={ \bDelta}_E/\bar{\tau}_E$: see Part II, (a)),
and it is possible to show that the Green-Kubo formula for
the conductivity $\bsigma\equiv\nabla_{ \bE}{ \bJ}|_{E=0}$
is equivalent to
$$
\nabla_{ \bE}\nu_E^{+}({\bold A}_E)|_{E=0}=\nu_0\left({\bold A}_0
     (-\nabla_{ E}H|_{E=0})\right),
$$
where $H$ is the formal Hamiltonian in Eq.(18). This 
is exactly the usual perturbation formula for the response of an equilibrium
system to a small change in the potential.

\noindent {\it (d) The van Kampen Argument Against Linear Response} 

In a paper published in 1971, ``The Case Against Linear Response Theory,''
[23] van Kampen made an argument that linear response theory of the form 
we have considered above could not be expected to apply except for
extremely tiny external fields, much smaller than those for 
which experiment exhibits its validity. Since linear response theory
works in practice very well, there is obviously a flaw in the van
Kampen argument rather than in the response theory! Therefore, the 
argument has received subsequently a great deal of critical
attention.

The basic point of the van Kampen argument was that one cannot expect
a {\it linear microscopic response} (i.e. of individual particle
trajectories) over macroscopic times like seconds, minutes, or hours
under an external perturbation, unless that perturbation is
exceedingly small. On this basis he challenged the theoretical foundation
of the microscopic response theory derivations of macroscopic linear 
transport laws. To demonstrate his point, van Kampen considered electrical 
conduction in a system of electrons which move freely except for occasional
collisions with impurities. An external field ${ \bE}$ then has the
effect of displacing the particle paths from their unperturbed positions 
over a time $t$ by at least an amount ${{1}\over{2}}t^2(e|{ \bE}|/m).$
van Kampen argued that, in order that the induced current be linear
in $|{ \bE}|,$ one must have ${{1}\over{2}}t^2(e|{ \bE}|/m)\ll d,$
where $d$ is a mean spacing of impurities. Obviously, the upper bound
on allowed field strengths $|{ \bE}|$ becomes more severe as $t$ 
increases. Taking $d\approx 100 \AA$ and $t$ to be a macroscopic time
of order 1 second or so, van Kampen found that the field must be 
less than $\approx 10^{-18} { V\,\,cm}^{-1}$!

We believe that the most basic problem with this argument comes 
from the fact that linear response theory deals with {\it probability
distributions} rather than individual phase space points. We agree
therefore with the discussion of Kubo et al. in [37]. (In fact, our
work can be regarded as a rigorous realization of the program proposed
there of ``stochastization'' of the system by an appropriate
coarse-graining in the phase space: our ``Markov sieve'' is a
sophisticated mathematical version of such a coarse-graining.) If one
examines the formal response theory calculations, for example which
lead to our Eq.(9) in Section (b), it is apparent that it is really
rigorous under relatively modest assumptions. The observation 
made in [13], Section 7.8, that the integrands in formulas like
Eq.(9) typically decay to nearly zero in a {\it microscopic} time,
makes it very reasonable that such uniform integrability as we 
required will hold. Therefore, the invariant measure of realistic 
systems have very likely a finite, linear correction such as in Eq.(9). 
Furthermore, for such a system the macroscopic current response will
be the sum of separate contributions of many individual electrons.
Therefore, one should expect that there will be a {\it law of large 
numbers} for the macroscopic current, so that for every phase space
point distributed with respect to the invariant measure the actual, 
empirical value of the current is equal to the average value with
a probability approaching unity as the number of electrons increases
(e.g. see [15]). Since the average value has a perfectly linear
behavior in $|{ \bE}|$, so will the actually observed current response.

Our simple model provides a strong counterexample to the
van Kampen argument, since instability of individual trajectories
holds in the strongest form (exponential separation in time), but
nevertheless the usual linear response theory is { \it rigorously} 
established. Hence we have the opportunity to examine in this
concrete situation the source of failure of van Kampen's argument.
In fact, it seems that the basic assertion on which the argument is 
constructed is actually false: in our model, {\it there is a linear
microscopic response over the whole infinite interval of time}!
This holds in a sense which is sometimes called the ``structural 
stability'' of hyperbolic flows to small perturbations. Perhaps
paradoxically, the dynamical systems which exhibit the strongest 
instability of individual trajectories in fact possess a strong
stability of the phase portrait as a whole. For smooth hyperbolic
systems, such as Axiom A flows, the precise statement of structural
stability is that, if $S_E^t$ is an $O(E)$ perturbation 
(in a suitable norm) of the flow $S^t$ on a manifold $M,$ then 
there is a diffeomorphism $J_E:M\rightarrow M,$
which moves points by a distance less than $E$ and such that
$J_E\circ S^t\circ (J_E)^{-1}=S_E^{ct}$
for some constant $c$, $|1-c|<E$ (the rescaling of time is just
to ensure that the isomorphic systems have the same entropy) [2].
In other words, for every phase point $X$ there is another nearby point
$X_E,$ within distance $E,$ such that $S^t(X)$
and $S_E^{ct}(X_E)$ are within $E$ of 
each other for {\it all} $t$. The structural stability in the above
form is not expected to hold for discontinuous systems like our
billiard model, but there is a slight weakening, termed ``stochastic
stability,'' which is conjectured to apply [32]. The statement of 
the property requires a stationary measure $\nu$ for $S^t,$ and is
intrinsically statistical. The difference
in the statement is that $J_E$ is now required only to
be measurable with a measurable inverse and to move points by
distance less than $E$ except possibly for a set of
$\nu$-measure less than $E.$ In other words, the points
$S^t(X)$ and $S_E^{ct}(X_E)$ are within $E$
of each other for all $t$ except for a fraction of the time proportional to 
$E.$ This is a much stronger form of stability than that observed numerically
in [31]---and also applied to discuss the van Kampen objection---which
corresponded merely to the linear proportionality of the trajectory
divergence (of the {\it same} phase point under the two dynamics) to the
applied field $E.$ However, the divergence was still exponential in
time, and after a lapse of time proportional to $\log(1/E)$ the two 
phase points are still well-separated in phase-space.
In contrast, we have shown here that, with a slightly more
flexible notion of stability, i.e. by choosing an appropriate
neighboring point rather than the {\it same} point for both dynamics,
the proportionality to $E$ is obtained uniformly for {\it all} times
(except for a negligible fraction of times, also proportional to $E.$)

Although the ``stochastic stability'' in the above precise form
is not proved here for our model, closely related properties are in 
fact used throughout our proofs. The entire hyperbolic structure of 
the billiards at zero field---local stable and unstable manifolds,
Markov partitions, etc.---are just ``slightly distorted'' by turning
on the thermostat and small field. Of course, we cannot say whether
similar stability properties will be present in realistic many-particle
situations, but the simple model gives some grounds for doubting 
the very basis of van Kampen's argument.

\newpage

\head 
Mathematical Proofs 
\endhead

Here we supply the proofs of all the main results in Part I.
First, we repeat in section (a) the formal response theory 
calculations in terms of the billiard map $T_E,$ but taking care to 
point out what is needed to make the argument into a proof.
Afterward, we turn to the more difficult
problem of existence of the limit measure $\nu^+_E$ and investigation
of its ergodic and statistical properties. Our reasonings
here are very much similar to those employed in the theory of hyperbolic
billiards [3-7,26,35]. Recently this theory has been extended
also to certain billiard-like Hamiltonian systems [12,8]
and to piecewise linear hyperbolic maps of the torus [38,9].
This theory is now sufficiently far developed, so that we
can only outline here the corresponding arguments. We will
explain in detail only the properties of our model which
differ from those of billiards and other related systems.
The main difference is certainly the absence of an absolutely continuous 
invariant measure in the phase space. In section (b) the existence of
local stable and unstable fibres for $T_E$ in $M$ (and for $S_E^t$
in $\frak M$) is established for almost every point with respect to 
the {\it zero-field} measures. In section (c) the main tools for
study of the statistical properties of the model, the so-called
{\it homogeneous fibres} and the {\it Markov sieve}, are introduced
and investigated. In the longest section (d) the stationary measure
$\nu_E^+$ is constructed and some basic probability estimates 
developed. In section (e) the uniform estimates on decay
of correlations are established which are needed to prove the
response formulas and Einstein relation. Finally, in section (f)
the Pesin and Young's formulas are established, as well as their
limiting behavior for small $E$ is investigated. 

The smallness of the field $E$ is always assumed as well as
the condition of finite horizon. 
Throughout the text we denote by $c_1,c_2,\ldots$ 
various positive constants (usually, constant factors),
by $a_1,a_2,\ldots$ also positive constants (usually,
exponents) and by $\alpha_1,\alpha_2,\ldots$ various
positive numbers which are less than $1$.

\noindent{\it (a) Response Calculation for the Discrete-Time Map}

We first make the calculation for the formula in Eq.(15)
of Proposition 3 which gives the expectations with respect to 
$\nu_E^+.$ Obviously, for any function $\phi$ on $M,$
$$\matrix
\hat{T}^n_E\nu_0(\phi)&=&\nu_0(\phi)
     +\sum_{k=1}^n[\nu_0(\phi\circ T_E^k)-\nu_0(\phi\circ T_E^{k-1})] \\
     &=&\nu_0(\phi)   
     +\sum_{k=1}^n\nu_0\left[(\phi\circ T_E^k)
     \left(1-{{d(\hat{T}_E^{-1}\nu_0)}\over{d\nu_0}}\right)\right].
\endmatrix
$$
To evaluate the Radon-Nikod\'{y}m derivative $d(\hat{T}_E^{-1}\nu_0)/d\nu_0,$ 
we recall first the obvious fact that
$$
{{d(\hat{S}_E^t\mu_0)}\over{d\mu_0}}
          =\left|{{\partial S_E^{-t}(X)}\over{\partial X}}\right| .
$$
The Jacobian determinant obeys the Eq.(6) in Part I which may
be explicitly integrated to give
$$
\left|{{\partial S_E^{-t}(X)}\over{\partial X}}\right|
    =\exp\left[-\beta{ \bE\cdot}\left({ \bQ}\circ S^{-t}_E
     -{ \bQ}\right)\right].
$$
For any subset $\Delta$ of $M$ and time interval $I$ we may define
the product set in the special flow coordinates $\Delta\times I
\equiv \cup_{\tau\in I}\,S_E^\tau[\Delta]$ contained in $\frak M.$
Since the particle moves at the same speed $v$ under $S_E^t$ for {\it each}
$E$, it follows that for a net $\{\Delta_x\}$ of Vitali sets converging 
to $x\in M,$ 
$$\matrix
{{d(\hat{T}_E^{-1}\nu_0)}\over{d\nu_0(x)}}
  &=&\lim_{\Delta_x\downarrow \{x\}}{{\nu_0(T_E[\Delta_x])}
     \over{\nu_0(\Delta_x)}} \\
  &=&\lim_{\Delta_x\downarrow \{x\}}\lim_{\delta T\downarrow 0}
     {{\mu_0\left(S_E^{\tau_E(x)}(\Delta_x\times [-\delta T,\delta T])\right)}
     \over{\mu_0(\Delta_x\times [-\delta T,\delta T])}} \\
  &=& {{d(\hat{S}_E^{-\tau_E(x)}\mu_0)}\over{d\mu_0(x)}}\,\,\,\,\,\,\,\,
\nu_0-a.e.\,\,x 
\endmatrix 
$$
Thus we obtain
$$\matrix
{{d(\hat{T}_E^{-1}\nu_0)}\over{d\nu_0}}&=&\exp\left[-\beta{ \bE\cdot}
    \left({ \bQ}\circ T_E -{ \bQ}\right)\right] \\
   &=&\exp[-\beta{ \bE\cdot\bDelta}_E]. 
\endmatrix 
\tag 19
$$
Observe from the above equality that $\nu_0(\exp[-\beta{ \bE\cdot\bDelta}_E])
=1.$ Therefore, finally 
$$
\hat{T}^n_E\nu_0(\phi)=\nu_0(\phi)
     +\sum_{k=1}^n\nu_0\left[(\phi\circ T_E^k)
     \left(1-e^{-\beta{ \bE\cdot\bDelta}_E})\right)\right].
$$
Now, if we assume that the measure $\hat{T}^n_E\nu_0$ converges weakly
as $n\rightarrow +\infty$ to $\nu_E^+$ and that the summation
converges also in that limit for at least $C^1$-smooth $\phi$, 
then we obtain exactly
$$
\nu_E^{+}(\phi)=\nu_0(\phi)+\sum_{n=1}^{\infty}\nu_0
     [(\phi\circ T_E^{n})(1-e^{-\beta { \bE\cdot\bDelta}_E})],
       \tag 20
$$
for such $\phi.$ If the convergence of the sum can be proven for
a larger class of functions by some uniform summable bound on its terms,
e.g. as below for $H_{\alpha}^*,$ then the formula can be extended to 
that class also by approximation.

We now provide the calculation for the second half of Proposition 3,
the nonlinear response formula. Since ${ \bv}=\dot{{ \bQ}}$, we 
see that
$$\matrix
{ \bJ}({ \bE})\equiv \mu_E^{+}({ \bv})&=&{{1}\over{\bar{\tau}_E}}\int_M 
    \nu_E^+(dx)\int_0^{\tau_E(x)}d\tau\,\,{ \bv}(x,\tau)  \\
    &=&{{1}\over{\bar{\tau}_E}}\nu_E^+({ \bQ}\circ T_E-{ \bQ}). 
\endmatrix 
$$
Thus, if Eq.(20) is shown to hold for $\phi={ \bDelta}_E/\bar{\tau}_E,$
we may simply substitute to obtain the formula for ${ \bJ}({ \bE}).$
However, it is somewhat more convenient to use the invariance to
write
$$
{ \bJ}({ \bE})={{1}\over{2\bar{\tau}_E}}\nu_E^+({ \bDelta}_E+
    \tilde{{ \bDelta}}_E),
$$
with $\tilde{{ \bDelta}}_E\equiv{ \bDelta}_E\circ T_E^{-1}=
{ \bQ}-{ \bQ}\circ T_E^{-1}.$
In that case, the contribution from the first term of Eq.(20)
is seen to vanish. Indeed, $\nu_0$ is invariant under time-inversion
and also ${ \bQ}(\tilde{x})={ \bQ}(x),$ so that $\nu_0({ \bQ}\circ T_E)
=\nu_0({ \bQ}\circ T_E^{-1}).$ On the other hand, the two terms from
the summation are easily calculated to give the response formula Eq.(16).

The next problem is to give arguments for the results on linear response in 
Proposition 4. We define a remainder function $R_E(x)\equiv 
\frac{\left(1-\beta{ \bE\cdot\bDelta}_0(x)
-e^{-\beta{ \bE\cdot\bDelta}_E(x)}\right)}{E},$
so that we may write
$$
\nu_E^{+}(\phi)=\nu_0(\phi)+\beta{ E\cdot}\sum_{n=1}^{\infty}
    \nu_0[(\phi\circ T_0^{n}){ \bDelta}_0]
    +E\cdot\sum_{n=1}^{+\infty}\nu_0[(\phi\circ T_0^{n})R_E]. \tag 21
$$
Observe that $\nu_0(R_E)=0$ for every ${ E}$. Furthermore,
$R_E$ is shown in section (e) to be bounded uniformly in $E$ and
$\lim_{E\rightarrow 0}R_E(x)=0\,\,\nu_0-a.e.$ (by the $C^2$ convergence 
of $T_E$ to $T_0$ for $E\rightarrow 0.$) From dominated convergence
the terms in the last sum of Eq.(21) go individually to zero.
Thus, only a summable bound on $\nu_0[(\phi\circ T_0^{n})R_E]$ uniform
in $E$ is required to infer that the last term is
rigorously $o(E).$ The argument for Ohm's law and the Einstein
relation is made similarly. Note first that ${ \bDelta}_E$ is 
bounded (uniformly in $E$) and $\lim_{E\rightarrow 0}\bDelta_E(x)
=\bDelta_0(x)\,\,\,\,\nu_0$-a.s. Thus, $\nu_0[{ \bDelta}_E(1-e^{-\beta {
\bE\cdot\bDelta}_E})-\beta({ \bDelta}_0\otimes{ \bDelta}_0)\cdot{ \bE}]=o(E).$
Likewise, $\nu_0[({ \bDelta}_E\circ T_E^n)(1-e^{-\beta 
{ \bE\cdot\bDelta}_E})-({ \bDelta}_0\circ T_0^n)\beta 
{ \bE\cdot\bDelta}_0]=o(E)$ for each $n$ and a suitable uniform decay bound 
gives also $o(E)$ for the summation. Thus, ${ \bJ}=\bsigma\cdot{ \bE}
+o(E)$ with 
$$
\bsigma=\beta\left(
           {{1}\over{2\bar{\tau}_0}}\nu_0({ \bDelta}_0\otimes{ \bDelta}_0)
           +{{1}\over{\bar{\tau}_0}}\sum_{n=1}^{+\infty}
           \nu_0(({ \bDelta}_0\circ T_0^n)\otimes{ \bDelta}_0)\right) ,
$$
which is obviously equivalent to what is stated in Eq.(17)
of Proposition 4.

\noindent{\it (b) Existence of Local Stable and Unstable Manifolds}

We need here a few additional notations. Set $S_n=T^{n-1}_ES_1$
and $S_{-n}=T^{-(n-1)}_E$\linebreak
$S_{-1}$ for every $n\geq 1$. A natural measure equivalent to the
length on a smooth curve $\gamma\in M$ is defined as
$$
    \rho (\gamma )= \int_{\gamma}\cos\varphi\, dr,
$$
in terms of the standard coordinates $(r,\phi),$
see [5,6,16].

\proclaim{Lemma 1 (Stable and unstable fibers)}
Almost every point $x\in M$ (with respect to the measure
$\nu_0$) has stable and unstable fibers, denoted by
$\gamma^s(x)$ and $\gamma^u(x),$ respectively, passing
through $x$.
\endproclaim 

   \demo{Proof} First we observe that the billiard system
$\{ S^t_0\}$ has here a finite horizon and a smooth strictly
concave boundary $\partial Q$. In particular, the time
of first return $\tau_0(x)$ is bounded away from $0$ and
$\infty$: $0<\tau_{\min}\leq\tau_0(x)\leq\tau_{\max}<\infty$.
These properties lead to a strong hyperbolicity of the
billiard map $T_0$. The hyperbolicity of $T$ can be defined in
terms of families of strictly invariant cones which are popular
nowadays, see e.g. [40,7,12]. These are two families of cones in
the tangent plane $\Cal T M$ to the manifold $M$ such that the
unstable cones are strictly invariant under $DT_0$ while the stable
ones are strictly invariant under $DT_0^{-1}$. Note that the rate of 
expansion (contraction) of each tanget vector in the unstable 
(stable) cones in our $\rho$-metric is bounded away from 1,
see e.g. [5,6]. We denote the minimal (maximal) rate of expansion 
(contraction) by $W_0>1$ (resp., $w_0<1$). Due to the smallness of $E$ 
the same cones are still invariant under $DT_E$ (respectively,
under $DT_E^{-1}$) and the minimal (maximal) rate of expansion 
(contraction) $W_E$ ($w_E$) are close to $W_0$ ($w_0$) and still
bounded away from 1. Nevertheless, we cannot apply here the
usual theorems on invariant cones [40] since no invariant
measure has been constructed yet for $T_E, E>0$. Instead,
we apply a direct method for constructing stable and
unstable fibers, see [34,35,5].

   For any point $x\in M$ and $n\geq 1$ we take a curve 
$\gamma_n^{\prime}$ passing through $T^{-n}x$ and lying in 
unstable cones. (This means that at every point of that curve 
the tangent vector to the curve belongs to the unstable cone.) 
Then $\lim_{n\to\infty}T^n\gamma^{\prime}_n$ gives us the unstable
fiber $\gamma^u_x$ provided that $T^n$ is continuous on $\gamma_n^{\prime}$
for all $n$ and the length of $T^n\gamma ^{\prime}_n$ is bounded 
away from $0$. To estimate that length one needs the bound
$$
      \nu_0(U_{\varepsilon}(S_{\pm 1}))\leq c_1\varepsilon
                                          \tag 22
$$
for all $\varepsilon >0$, where $U_{\varepsilon}(S_{\pm 1})$
denotes the $\varepsilon$-neighborhood of $S_{\pm 1}$.
The estimate (22) readily comes from the fact that in case
of finite horizon the set $S_{\pm 1}$ is always a finite
union of smooth compact curves in $M$. Standard arguments
[34,35,5] show that $\gamma^u(x)$ has the $\rho$-length
$\geq\varepsilon_0$ as soon as $T^{-n}x$ lies outside
$U_{\varepsilon_n}(S_{\pm 1}), \varepsilon_n=\varepsilon_0
\cdot w_E^n$ for all $n\geq 1$. Therefore, the set
of points $x\in M$ with the unstable fiber $\gamma^u(x)$
of length $<\varepsilon_0$ has the $\nu_0$-measure less than
$$
      \sum_{n=0}^{\infty}\nu_0(T^n_EU_{\varepsilon_n}(S_{\pm 1})).
                                     \tag 23
$$
The map $T_E$ can contract or expand the measure $\nu_0$
but with a rate not greater than $\exp(\omega E)$
where $\omega=p\beta\tau_{\max}/m$ is a constant,
see the Section (a) and [30]. For small $E$ this rate is close to $1$ and
therefore less than $w_E^{-1}$. The sum (23) then
does not exceed
$$
    c_2\varepsilon_0\sum_{n=1}^{\infty}
          (\exp(\omega E))^{-n}w_E^n
                =c_3\varepsilon_0 .
$$
Hence the lemma. \qed \enddemo

\proclaim{Corollary 2} The $\nu_0$-measure of the set of points
$x\in M$ for which the $\rho$-length of the unstable (stable) 
fiber is less than $\varepsilon$ does not exceed $c_4\cdot\varepsilon$.
\endproclaim

{\smc Remark}. In the proof of Lemma 1 we have explicitly 
found the necessary bound on $E$: $\exp[Ep\beta\tau_{\max}/m]
<w_E^{-1}$. However, in our further considerations
we can no longer do so.

\proclaim{Corollary 3} For almost every point $X$ in the
whole space $\frak M$ (with respect to the measure $\mu_0$)
there are stable and unstable manifolds of the flow
$\{S^t_E\}$ passing through $X$.
\endproclaim

  \demo{Proof} Due to Lemma 1 there is a stable curve
$\gamma^s(x)$ for $\nu_0$-almost every $x\in M$. It certainly
provides a bunch of trajectories which converges
exponentially fast in the future. Next, for any point
$y\in\gamma^s(x)$ denote $\theta_n(y)=\tau(y)+\tau(T_Ey)
+\cdots +\tau(T_E^{n-1}y)$ the time up to the $n$th
reflection. It is now clear that $|\theta_n(y)-
\theta_n(x)|\leq\sum_0^{n-1}|\tau(T_E^iy)-\tau(T_E^ix)|
\leq c_5\cdot dist(x,y)$. Therefore, the function
$\Delta(y)=\lim_{n\to\infty}(\theta_n(y)-\theta_n(x))$
is continuous on $\gamma^s(x)$ and has a finite derivative
at $x$ with respect to the $\rho$-length. Now the map
$y\mapsto S^{\Delta(y)}y$ transforms the fiber $\gamma^s(x)$
into the stable manifold for the flow $\{S^t_E\}$. The
projection of that manifold into $Q$ is a curve transversal
to the trajectories of the flow $\{S^t_E\}$. Note that
generally that curve is not orthogonal to the trajectories
of the flow, as it was in case of billiards. \qed \enddemo

   In the case of billiard flow $\{S^t_0\}$ the curvature of a
stable (unstable) manifold at a point $X\in\frak M$ is expressed
through a continued fraction $B^s(X)$ (resp., $B^u(X)$), see [34,36].
The differential equation of the stable (unstable) fibers
in $M$ is then readily obtained as
$$
     \frac{d\varphi}{dr}=-B^s(x)\cos\varphi+\varkappa(x)\;\;\;\;\;
     (\frac{d\varphi}{dr}=B^u(x)\cos\varphi+\varkappa(x))
        \tag 24
$$
where $\varkappa(x)$ stands for the (positive) curvature of
the boundary $\partial Q$ at the point $x$. 
   In our model, with $E>0$, the curvature of stable or
unstable manifold is no longer expressed through any
continued fraction. But, if we denote by $B^s_E(X)$
($B^u_E(X)$) the curvature of the orthogonal section
of the beam of trajectories generated by the stable
(unstable) manifold at $X\in\frak M$, then the expressions
(24) remain true for $E>0$.

\proclaim{Lemma 4 (Absolute continuity)}
The stable and unstable fibers in the space $M$ are 
absolutely continuous with respect to the measure $\nu_0$.
\endproclaim

   The statement of Lemma 4 means that the canonical
isomorphism on stable and unstable fibers, see eg. [3,6],
is absolutely continuous with respect to the $\rho$-length
on those fibers. The proof of Lemma 4 goes the same way
as that of its analogue for billiards [34,16] and we do
not go into detail.

\proclaim{Remark 5 (Alignment)}
The images $S_n$ of the singularity curves lie in unstable cones
for $n>0$ and in stable cones for $n<0$. Thus, they become
almost parallel to unstable fibers as $n\to\infty$ and to
stable fibers as $n\to -\infty$.
\endproclaim

We do not make this statement more precise.

\noindent{\it (c) Homogeneous Fibres and Markov Sieves}

   Stable and unstable curves with the absolute continuity
property constitute the main tool for the study of ergodic
properties of hyperbolic systems.  But the exploration
of their statistical properties requires the so called {\it 
homogeneous fibers}. These fibers have been first introduced for
billiards in [6]. As it was explained there, the billiard
map $T_0$ expands unstable manifolds but nonuniformly: the
rate of expansion grows in the neighborhood of $\partial M$
where $\cos\varphi$ vanishes. In order to control this
rate the authors of [6] splitted the neighborhood of $\partial M$
into a countable number of strips the thinner the closer
to $\partial M$. The strips were defined by the equations
$\pi/2-(n+1)^{-\eta}\geq\varphi\geq\pi/2-n^{-\eta}$
in the neighborhood of the line $\varphi=\pi/2$ and 
$-\pi/2+(n+1)^{-\eta}\leq\varphi\leq-\pi/2+n^{-\eta}$
in the neighborhood of the line $\varphi=-\pi/2$, where
$n\geq n_0$. The parameters $\eta >1$ and $n_0\geq 1$
are rather arbitrary except $n_0$ should be large enough.
We denote $\frak D_0$ the union of the lines separating the strips.

  {\smc Definition.} An unstable (stable) fiber $\gamma^u$
($\gamma^s$) is said to be {\it homogeneous} (or
{\it $0$-homogeneous}) if its
images $T^{-n}\gamma^u$ ($T^n\gamma^s$) for $n\geq 0$
never cross $\frak D_0$,
the borders of the above strips. An unstable
(stable) fiber $\gamma^u$ ($\gamma^s$) is said to be
$m$-homogeneous, $m\geq 1$, if its larger preimage
$T^m\gamma^u$ ($T^{-m}\gamma^s$) is a homogeneous
fiber.

   The following lemmas have been proved in full detail
for billiards in [6]. For small $E$ they are also valid for 
our system and the proofs are essentially the same.

\proclaim{Lemma 6 (Existence)} Almost every point $x\in M$
(with respect to $\nu_0$) has homogeneous stable and
unstable fibers passing through $x$.
\endproclaim

   The largest smooth components of the homogeneous
stable and unstable fibers passing through $x$ are
denoted $\gamma^{0s}(x)$ and $\gamma^{0u}(x)$ respectively.
The next lemma is a natural extension to Corollary 2.

\proclaim{Lemma 7 (Distribution of length)} For every
$\varepsilon>0$ the set of points $x\in M$ with the
homogeneous fibers of length $<\varepsilon$ has 
$\nu_0$-measure less than $c_6\varepsilon^{a_1}$,
where $a_1$ depends on the choice of the value of $\eta$
above.
\endproclaim

   For every point $x\in M$ and $k\geq 1$ denote by
$w^u_k(x)$ the rate of contraction of $\gamma^u(x)$
at the point $x$ under $T^{-k}$.

\proclaim{Lemma 8 (Homogeneity)} Let $\gamma^{0u}$ be
an arbitrary $m$-homogeneous unstable fiber, $m\geq 0$. Then
for  every pair $x,y\in\gamma^{0u}$ and every $k\geq 1$
$$
    \left |\frac{w^u_k(x)}{w^u_k(y)}-1\right |
           \leq c_7\alpha_1^m ,
$$
where $c_7,\alpha_1$ are determined by $\eta$ and $n_0$ above.
\endproclaim

   Our further considerations extensively use elements of Markov
partitions for hyperbolic systems. In our notions and
notations we follow the traditions of works [3,5,6].
A basic notion in the theory of Markov partitions is
a {\it parallelogram\/}. It is defined as a subset $A\subset M$
such that for any two points $x,y\in A$ the point
$z=\gamma^u(x)\cap\gamma^s(y)$ exists and again belongs to $A$.
If we substitute $\gamma^{0u}(x)$ and $\gamma^{0s}(y)$ for
$\gamma^u(x)$ and $\gamma^s(y)$ in this definition, we
obtain the definition of
a {\it homogeneous parallelogram}. If for a parallelogram
$A$ both its images $T^mA$ and $T^{-m}A$ are homogeneous
parallelograms, then $A$ is said to be {\it $m$-homogeneous}.
In what follows we always consider only homogeneous
parallelograms without specifying this.

  Any parallelogram $A$ is a Cantor set with a grid structure.
We denote $\gamma^{u,s}_A(x)=\gamma^{u,s}(x)\cap A$ for every
$x\in A$. The sets $\gamma^u_A(x)$ (and $\gamma^s_A(x)$) for all
$x\in A$ are Cantor sets on the corresponding fibers which
are canonically isomorphic, see e.g. [6].

   Let $A_0$ be an $m$-homogeneous parallelogram and $x_0\in A_0$.
As shown in [6], the $\nu_0$-measure of any subparallelogram 
$A\subset A_0$ can be approximated by the value
$$
        \nu_0^a(A)=\rho(\Gamma^u_A)\rho(\Gamma^s_A)
                     (B^u(x_0)+B^s(x_0)).   \tag 25
$$
Here $\Gamma^{u,s}_A$ denote the images of $A$ on the
fibers $\gamma^{u,s}(x_0)$ under the canonical isomorphisms.
More precisely, the value (25) is an approximation to the 
$\nu_0$-measure of $A,$ constructed below, with an exponentially 
small error:
$$
   |\nu_0^a(A)/\nu_0(A)-1|\leq c_8\alpha_2^m.   \tag 26
$$ 

  Evidently, the image $T_E^nA$ of a parallelogram $A$ is a
finite union of parallelograms. Consequently, the intersection 
$T_E^nA\cap B$ is again a finite union of parallelograms, where 
$B$ is another parallelogram. We say that a subparallelogram 
$C\subset B$ is {\it $u$-inscribed} ({\it $s$-inscribed}) in $B$ 
if $\gamma^u_C(x)=\gamma^u_B(x)$ (resp. $\gamma^s_C(x)=\gamma^s_B(x)$) 
for every $x\in C$. Of the parallelograms composing $T_E^nA\cap B,$ 
the union of those $u$-inscribed in $B$ and such that their images 
under $T_E^{-n}$ are $s$-inscribed in $A$ is called the {\it regular 
part} and denoted by ${\frak R}(T_E^nA\cap B)$, while the union of 
the others is called the {\it irregular part} of that intersection 
and denoted by ${\frak I}(T_E^nA\cap B)$. Dual notations are introduced 
for $T_E^nA\cap B$ with $n\leq -1$. The intersection $T_E^nA\cap B$ 
is said to be {\it regular} if it contains no irregular part.

   The {\it Markov partition\/} for $T_E$ is a countable partition 
(mod $0$) of the manifold $M$ into parallelograms $\{A_1,A_2,\ldots\}$ 
such that the intersections $T^nA_i \cap A_j$ are regular for any pair 
$A_i,A_j$ and any $n\neq 0$. Note that there cannot be finite Markov 
partitions due to the presence of arbitrary short fibers. Markov 
partitions for $T_0$ have been constructed in [3,5]. This construction 
can be extended to $T_E$ with small $E>0$. However, it seems to be of 
no use for us because it is not clear whether the parallelograms of 
the Markov partition cover a.e. point in $M$ with respect to the needed
measure $\nu^+_E$.

   We use {\it Markov sieves} introduced in [6,9] and defined below. 
They consist of a finite number of parallelograms and therefore do 
not cover a set of full measure in $M$. But the Markov sieve turns 
out to be much easier to construct and to control than the Markov 
partition, and it also yields useful estimates of the statistical 
properties of hyperbolic dynamical systems with singularities [6,9]. 
Let us stress also that the Markov sieves depend on the interval of
time which is considered.
 
  The Markov sieves are closely related to the {\it pre-Markov
partitions} [5,6,9] and we define them both below.

   Any domain $\Pi$ in $M$ bounded by two unstable and two stable 
fibers is called the {\it quadrilateral}. Its boundary $\partial \Pi$ 
consists of two unstable fibers called the {\it $u$-sides of $\Pi$} 
and two stable ones called the {\it $s$-sides of $\Pi$}. The union
of two $u$-sides is denoted by $\partial^u\Pi$ and that of two $s$-sides 
is denoted by $\partial^s\Pi$. Fix a sufficiently large $m\geq 1$ and 
let $\varepsilon>0$ be arbitrarily small and real ($\varepsilon<
\varepsilon_0(m)$). A {\it pre-Markov partition} for the map $T^m$ is 
a finite partition $\xi_0=\xi_0(\varepsilon)$ of $M$ into curvilinear
polygons $P_1,\ldots,P_k$ whose properties are listed next: The boundary 
$\partial\xi_0=\cup\partial P_i$ is the union of $S_{-m,m}=\cup_{k=-m}^m\,
\,S_k$ and a finite collection of unstable and stable fibers. Respectively, 
we denote $\partial\xi_0=\partial^0\xi_0 \cup\partial^u\xi_0\cup\partial^s
\xi_0$, where $\partial^0\xi_0=S_{-m,m}$ and $\partial^u\xi_0$ ($\partial^s
\xi_0$) consists of unstable (stable) fibers. The main properties of 
$\xi_0$ are $T(\partial^s\xi_0)\subseteq\partial^s\xi_0$ and $T^{-1}
(\partial^u\xi_0)\subseteq\partial^u\xi_0$. All the interior angles of 
the polygons $P\in\xi_0$ both sides of which are unstable and stable 
fibers are always less than $\pi$. The sides of the polygons $P\in\xi_0$ 
lying on unstable (stable) fibers are less than $c_9\varepsilon$ and 
their images under $T^m$ (resp., $T^{-m}$) remain less than $c_{10}
\varepsilon$. If a polygon $P_i\in\xi_0$ does not touch the set $S_{-m,m}$, 
then it is a quadrilateral. All the other elements of $\xi_0$ form a 
neighborhood of $S_{-m,m}$ which we call the {\it necklace} and denote 
by $\frak N (\xi_0)$. The necklace is actually contained in a $c_{11}
\sqrt{\varepsilon}$-neighborhood of $S_{-m,m}$, and so its $\nu_0$-measure 
is less than $c_{12}\varepsilon^{a_2}$. We also define an {\it extended
necklace\/} $\frak N_e(\xi_0)$ as the union of $\frak N(\xi_0)$ and all 
the quadrilaterals $\Pi\in\xi_0$ intersecting $\frak D_0$, the borders 
of the strips constructed in the definition of homogeneous fibers. It 
is easily checked that $\nu_0(\frak N_e(\xi_0))\leq c_{13}\varepsilon^{a_3}$.

\proclaim{Remark 9} Every stable and unstable fiber is
either transversal to $S_{-m,m}$ or tangent to it,
and in the latter case the tangency has the order two,
see [34,3]. Therefore, the necklace $\frak N(\xi_0)$
can cover only a small part of that fiber so that the
total $\rho$-length of that part is less than $c_{14}
\varepsilon^{a_4}$. The extended necklace $\frak N_e(\xi_0)$ 
also has that property. Likewise, the $\varepsilon$-neighborhood 
of $S_{-m,m}$ for any $\varepsilon>0$ covers only a small part 
of that fiber so that the total length of that part is less than $c_{15}
\varepsilon^{a_5}$.
\endproclaim

   For precise description of the evolution of
parallelograms in $M$ we use the following geometrical
notions introduced in [6,3].
For any parallelogram $A$ the minimal
closed quadrilateral containing $A$ is called
the {\it support} of $A$ and denoted by $\Pi(A)$.
We say that a segment of an unstable (stable) fiber
is {\it inscribed} in a quadrilateral $\Pi$ if it lies within $\Pi$
and terminates on two $s$-sides ($u$-sides) of $\Pi$.
A parallelogram $A$ is said to be {\it maximal}
if it intersects all the unstable and 
stable fibers inscribed in its support $\Pi(A)$.
In other words, to construct a maximal parallelogram
one should take a quadrilateral $\Pi$, draw all the unstable
and stable fibers inscribed in $\Pi$ and take all the
mutual intersection; thus the maximal parallelogram
would consist of the points of intersections of these 
fibers. The parallelogram so obtained is denoted by $A(\Pi)$.

   Now let $n\geq 1$ be a large number and $\varepsilon_n
=\alpha_3^n$ for some $\alpha_3\in (0,1)$. Consider
the partition $\xi_n=T^{-n}\xi_0\vee T^{-n+1}\xi_0
\vee\ldots\vee T^n\xi_0$ of the space $M$, where
$\xi_0=\xi_0(\varepsilon_n)$ is a pre-Markov partition.
Denote $\frak N_e(\xi_n)=T^{-n}\frak N_e(\xi_0)\cup
\ldots\cup T^n\frak N_e(\xi_0)$. Clearly, $\nu_0
(\frak N_e(\xi_n))\leq c_{16}\alpha_4^n$ for some $\alpha_4
\in (0,1)$. Every element $\Pi$ of $\xi_n$ which
lies outside $\frak N_e(\xi_n)$ is a quadrilateral.
Moreover, its images $T^i\Pi$ for $|i|\leq n$ do
not intersect $S_{-m,m}$ or $\frak D_0$. Let
$\Pi_1,\ldots,\Pi_I$ be all the elements of $\xi_n$
lying outside $\frak N(\xi_n)$. The maximal
parallelograms $A_1=A(\Pi_1),\ldots,A_I=A(\Pi_I)$
form the {\it Markov sieve} which we denote
$\frak S_n$. The properties of the partition $\xi_n$
and Lemma 7 ensure that $\nu_0(M\setminus\cup A_i)
\leq c_{17}\alpha_5^n$ for some $\alpha_5\in (0,1)$. All
the parallelograms $A\in\frak S_n$ are maximal
and $n$-homogeneous. Note that if an unstable
fiber $\gamma_1^u$ crosses both $s$-sides of
an element $A\in\frak S_n$, then it intersects
$A$ itself. If a fiber $\gamma_1^u$ intersects
no parallelograms $A\in\frak S_n$, then it is
either too short (i.e. $\rho(\gamma_1^u)\leq
c_{18}\alpha_6^n$ for some $\alpha_6\in (0,1)$) or
it lies mostly in $\frak N_e(\xi_n)$. In the
latter case one of its images $T^i\gamma_1^u$
for some $|i|\leq n$ belongs to the extended
necklace $\frak N_e(\xi_0)$.

\noindent{\it (d) Existence of the Invariant Measure}

We now turn to the construction of the limit measure
$\nu^+_E$ which is defined as the limit of $\hat{T}^n_E\nu_0$
as $n\to\infty$. The conditional measure induced by $\nu_0$
on a segment of an unstable fiber $\gamma^u$ can be
constructed as follows. For $n\geq 1$ take a uniform probabilistic
measure (with respect to the $\rho$-length) on the
preimage $T^{-n}\gamma^u$ and then pull it back onto
$\gamma^u$. The limit of the resulting measure on $\gamma^u$
as $n\to\infty$ gives the conditional measure on $\gamma^u$.
Lemma 8 assures that the density of
the conditional measure on any homogeneous unstable
fiber is uniformly bounded away from $0$ and $\infty$.

   Let $\gamma^u$ be a homogeneous unstable fiber and $p^c$ 
denotes the conditional absolutely continuous probabilisty 
measure on $\gamma^u$ constructed above. It is now clear that 
the existence of the measure $\nu^+_E$ is equivalent to the fact 
that the limit of $p^c_n=\hat{T}_E^np^c$ as $n\to\infty$ exists 
and is independent of $\gamma^u$. This limit thus produces
the measure $\nu^+_E$ itself. Note that the methods of
[42] can give a weaker result, i.e. the existence
of $\lim_{n\to\infty}n^{-1}(\nu_0+\hat{T}_E\nu_0+\cdots
+\hat{T}_E^{n-1}\nu_0)$.

  The measure $p^c_n$ for finite $n$ is concentrated
on the image $T^n\gamma^u$ which is a finite or countable
union of homogeneous unstable fibers. These fibers are called 
{\it homogeneous components} of $T^n\gamma^u$, see [6], or just
{\it components}, for brevity. The structure and the
distribution of those components in the space $M$ play the
key role in our further considerations. The necessary
properties of the components are accumulated in the next
several lemmas. These lemmas have been first established 
for billiards in [6] and then for piecewise linear
toral maps in [9].

   For any $D>0$ and $n\geq 1$ denote $\Gamma_{n,D}^u$
the union of all components of $T^n\gamma^u$ which have 
$\rho$-length $\geq D$.

\proclaim{Lemma 10 (From short to long components)}
There is $D>0$ not depending on $n$ such that for any $n\geq 1$
$$
    p^c(\gamma^u\setminus\cup_{k=1}^nT^{-k}
        \Gamma_{k,D}^u)\leq c_{19}\alpha_7^n/\rho(\gamma^u)
$$
with some $c_{19},\alpha_7$ determined by $\eta$ and $n_0$.
\endproclaim

   The meaning of the lemma is that during the first $n$
iterates of $T$, if $n\geq -c_{20}\ln\rho (\gamma^u)$,
the majority of points $x\in\gamma^u$
appear at least once in long components (of length $\geq D$)
of the images $T^k\gamma^u, 1\leq k\leq n$. 

   The proof of Lemma 10 is based solely on the hyperbolic
properties of the underlying map. It has been carried out
in detail in [6,9] and applies to our system, too.

\proclaim{Lemma 11 (Distribution of lengths of components)}
For any $\varepsilon >0$ and $n\geq -c_{21}\ln\rho(\gamma^u)$ we 
have $p^c_n(\Gamma_{n,\varepsilon}^u)\geq 1-c_{22}\varepsilon$.
\endproclaim

\demo{Proof} Lemma 11 is just a stronger version of Lemma 10,
but it is new and so we outline its proof here.
The billiard map possesses the following basic property: for
every $m\geq 1$ the number of smooth components of
$S_{-m,m}$ meeting at a single point of $M$ cannot
exceed $K_0m$, where $K_0$ is a constant, see [5], sect.8.
As a result for every $m\geq 1$ there is $\varepsilon_0(m)>0$
such that any unstable fiber of length $\leq
\varepsilon_0(m)$ can cross at most $K_0m$ curves
of $S_{-m,0}$. This property is certainly valid
for the map $T_E$ for small $E, E<E_0(m)$. Now
we fix $m$ sufficiently large, so that $W^m_E
\gg K_0m$. Thus the image $T_E^m\gamma^u_1$ of any
short fiber $\gamma^u_1$ of length $\leq\varepsilon_0(m)$
consists of at most $K_0m+1$ components and their total
length is at least $\Lambda^m_E$ times greater than
that of $\gamma^u_1$. Similar estimates can be carried
out for homogeneous components, i.e. if we take into
account the splitting of the components by the 
borderlines of the strips defined above. The technique
used for obtaining those estimates is the same 
as in proof of Lemma 7, see [6] for details. Now we
introduce a function $r_n(x)$ on $T_E^n\gamma^u$
by $r_n(x)=\{\rho$-distance from $x$ to the nearest
endpoint of the component of $T_E^n\gamma^u$
containing $x\}$. Note that $r_n(x)$ is actually {\it smaller}
than the length of the component of $T_E^n\gamma^u$,
containing $x$. The above reasonings show that the
distribution of $r_n(x)$ cannot concentrate near
$0$, i.e. $p^c_n\{r_n(x)<\varepsilon\}\leq
c_{23}\varepsilon$. Hence the lemma.\qed\enddemo

   Since we have not yet proved the existence of $\nu^+_E$,
we denote by $\nu^U_E(B)$ and $\nu^L_E(B)$ the upper and
lower limits, respectively, of the sequence $\{p_n^c
(B)\}^{\infty}_0$ for any measurable set $B$. The values
$\nu_E^{U,L}(B)$ may also depend on the choice of the
initial fiber $\gamma^u$. 

   As an immediate consequence of Lemma 11
we obtain that the limit measure $\nu^+_E$ (if it exists)
is nonatomic. In terms of $\nu_E^{U,L}$ this means
that $\lim_{\delta\to 0} \nu_E^U(V_{\delta}(x))=0$
for any point $x\in M$, $V_{\delta}(x)$ being here
the $\delta$-disc centered at $x$. The following
remark is a stronger version of this property:

\proclaim{Remark 12 (Nonatomic structure)}
For any unstable fiber $\tilde{\gamma}^u$ we have
$\lim_{\delta\to 0}$\linebreak
$\nu^U_E(V_{\delta}(\tilde{\gamma}^u))=0$,
where $V_{\delta}(\cdot)$ now denotes the $\delta$-neighborhood.
\endproclaim

   Next, let $\gamma^u$ be an unstable fiber of length
$D/2$ and $\Pi$ be a quadrilateral in $M$. Denote $\Gamma
_{n,\Pi}^u$ the union of all subfibers in $T_E^n(\gamma^u)$
which are $u$-inscribed in $\Pi$.

\proclaim{Lemma 13 (From long components into a fixed quadrilateral)}
There exists a quadrilateral $\Pi$ such that $\nu_0(A(\Pi))>0$
and constants $n_1>0, \beta_1\in (0,1)$ such that
$p_n^c(\Gamma_{n,\Pi}^u)\geq\beta_1$ for every $n\geq n_1$.
Here $n_1$ and $\beta_1$ are independent of $\gamma^u$
and of the field $E$.
\endproclaim

   \demo{Proof} The proof of Lemma 13 for billiards is based
on the mixing property. Here we do not have it, so the
arguments should be modified. For billiard map $T_0$ the
statement of Lemma 13 has been proven [6] for any quadrilateral
$\Pi$ such that $\nu_0(A(\Pi))>0$ and any unstable fiber $\gamma^u$
of length $\geq D$. The proof is easily modified if, instead
of the fiber $\gamma^u$, we take any curve $\tilde{\gamma}^u$
of length $\geq D$ which is sufficiently close to unstable
fibers (to be specific, such that $T_0^{-m}$ is smooth on
$\tilde{\gamma}^u$ and $T_0^{-m} \tilde{\gamma}^u$ lies in
unstable cones, $m$ being a large constant). Now we take an
unstable fiber $\gamma^u$ of length $\geq D$ for $T_E, E>0$.
For small $E$, a bit smaller part of $\gamma^u$ (of length
$\geq D-\varepsilon$) is certainly a curve close to unstable
fibers in the above sense. The image $T_E^n\gamma^u$ is
close to $T_0^n\gamma^u$ for all $n\leq n_2$ due to the smallness
of $E$, and $n_2$ here is large for small $E$. Therefore
the statement of lemma follows for all $n, n_1\leq n\leq n_2$,
with maybe smaller values of $D$ and $\beta_1$ than in the case
of the billiard map $T_0$. To prove Lemma 13 for larger values
of $n$, i.e. for $n\geq n_2$, we observe that due to Lemmas 10, 11 
there are enough components of length $\geq D$ in the images
$T_E^n\gamma^u$ for $1\leq n\leq n_2-n_1$. To each of
those components we apply the above reasonings again, etc.
Thus we extend our estimate for all $n\geq n_1$.\qed\enddemo

\proclaim{Corollary 14}
$\nu^L_E(\Pi)>const>0$ for any initial fiber $\gamma^u$.
Note also that each fiber $\gamma^u_1$ $u$-inscribed 
in $\Pi$ intersects the parallelogram $A=A(\Pi)$ and
$\rho(\gamma^u_1\cap A)\geq C(A)\cdot\rho(\gamma^u_1)
\cdot\nu_0(A)$, due to the homogeneity of the parallelogram
$A$. Therefore, we also have $\nu^L_E(A)\geq C_2(A)>0$.
\endproclaim

   {\smc Remark}. The proof of Lemma 13 requires the
quadrilateral $\Pi$ to be small enough and $\nu_0(A(\Pi))>0$.
However, we {\it cannot} state Lemma 13 
and Corollary 14 for all quadrilaterals
with such properties. Indeed, we have supposed the field $E$
to be small enough {\it after} choosing $\Pi$, i.e. actually
we have required $E<E_0(\Pi)$. 

   Let $A=A(\Pi)$ be a maximal parallelogram with the support
$\Pi$ involved in Lemma~13. Consider a new map $T_{\ast}$ 
defined only on the fiber $\gamma^u$ and on the components
of its images. This map is specified by an ``absorbing" property 
of the parallelogram $A=A(\Pi)$. It acts exactly as the map
$T_E$ unless a component, $\gamma^u_1$, intersects both $s$-sides 
of the quadrilateral $\Pi$. In that last case the part $\gamma^u_1
\cap A$ stops moving under $T_{\ast}$, and then all the future
iterates of $T_{\ast}$ on that part are identities. The remaining
part, i.e. $\gamma^u_1\setminus A$, consists of a countable number
of curves -- subcomponents -- on which $T_{\ast}$ still acts as
the map $T_E$. After $n\geq 1$ iterates of $T_{\ast}$, a part
of $\gamma^u$ will be sooner or later ``stuck" with the parallelogram 
$A$ while the remaining part of it will be still moving. We
denote that remaining part by $\tilde{\gamma}^u(n)$. Obviously, 
its $p^c_n$-measure monotonically decreases in $n$.

   The next lemma is a natural extension of Lemma~13.
It was first introduced in [9] for piecewise linear toral
maps. Its proof [9] is based on the hyperbolic properties
of $T$ alone, so it works in our situation as well.

\proclaim{Lemma 15 (From long components into a fixed parallelogram)}
   For any fiber $\gamma^u$ of length $\geq D$ and any
$n\geq 1$ one has $p^c(\tilde{\gamma}^u(n))\leq c_{24}\alpha_8^n$
where $c_{24}>0$ and $\alpha_8\in (0,1)$ are constants, both 
independent of $\gamma^u$.
\endproclaim

   Roughly speaking, Lemmas 10, 11 say that a short fiber
is sufficiently fast transformed into long fibers, Lemma~13
says that a long fiber sufficiently fast sends some of its 
portions into fibers $u$-inscribed in $\Pi,$ and Lemma~15 tells 
that a long fiber is sufficiently fast transformed under $T_{\ast}$
into Cantor sets lying on fibers $u$-inscribed in $\Pi$ and covering
the points of $A(\Pi)$ on those fibers.

   {\smc Remark}. Our dynamics is obviously reversible.
That is, our Lemmas 10-15 have dual forms for stable
fibers and negative powers of $T_E$ (or, respectively,
the iterates of a new map $T^{(-)}_{\ast}$ which can be
defined in a similar fashion as $T_{\ast}$ by the action
of $T_E^{-1}$ and an ``absorbing'' property of $A$).

   The last remark actually provides the tool for the estimation
of the values of  $\nu_E^{U,L}(B)$ for arbitrary 
parallelogram $B$. First we describe the main idea of that
estimation and then work out the details. Let $B$ be an 
arbitrary small parallelogram with $\nu_0(B)>0$, which is 
also homogeneous and maximal. Consider an arbitrary $y\in B.$
By the dual statements to Lemmas 10-15 the fiber $\gamma^s(y)$
is sufficiently fast transformed into $A$ under
$T^{(-)n}_{\ast}, n\geq 1$, see the above remark. Each 
time when a component of $T^{(-)n}_{\ast}\gamma^s(y)$ 
crosses both $u$-sides of $A$, it also intersects $A$ 
and the points of $T^{-n}B$ on that component cover all 
the points of $A$ on it, due to maximality of both $A$ and $B$. 
We can extend the definition of $T^{(-)}_{\ast}$ to the whole
parallelogram $B$ and its images under $T^{-n}, n\geq 1$.
This means that $T^{(-)}_{\ast}$ acts exactly as $T^{-1}_E$
unless a component of $T^{-n}_E B$ crosses both $u$-sides
of $\Pi(A)$. In that last case the intersection $T^{-n}_E B
\cap A$ stops, and on the remaining part,  $T^{-n}_E B\setminus A$,
which consists of a countable number of parallelograms -- 
subcomponents -- the map $T^{(-)}_{\ast}$ will still act as $T^{-1}_E$.
Thus, by the lemmas 10-15, the parallelogram $B$ itself is 
sufficiently fast transformed into $A$ under $T^{(-)}_{\ast}$.

Fix now a large $n_0=n_0(B)$ and consider the sets $B^{(-)}_n=
T_{\ast}^{(-)(n-n_0)}(T_E^{-n_0}B$. In other words,
$B^{(-)}_n, n\geq 1$ are produced by the evolution of $B$
in the past when during the first $n_0$ iterates only it
evolves ``freely'' under $T^{-1}$ and then the ``absorbing''
property of $A$ is turned on. The sets $B^{(-)}_n$ for large 
$n>n_0$ then consist of a finite number of subparallelograms 
``stuck'' with $A$ and of some parallelograms outside $A$
which are still moving under $T_{\ast}^{(-)}$. The preimages 
of the former are disjoint subparallelograms in $B$. Clearly,
each of those preimages is transformed into $A$ by $T^E_{-k}$ 
for some $k>n_0$ and then its image under $T_E^{-k}$ belongs 
either to $\frak R(T^{-k}B\cap A)$ or to $\frak I(T^{-k}B\cap A)$.
We denote all the parallelograms of the first (regular) kind
by $B_1,B_2,\ldots$. Each $B_i,i\geq 1$ is a subparallelogram
in $B$ and there is a $k_i>n_0$ such that $T^{-k_i}B_i\subset
\frak R(A\cap T^{-k_i}B)$. The parallelograms of the second 
(irregular) kind are less important for us and we denote their 
union by $B^{(0)}$.

   Now we can estimate the measure $p^c_N(B)$
for large values of $N$. For each $i\geq 1$
the parallelogram $T^{-k_i}B_i$ is $s$-inscribed
in $A$, so that $p^c_{N-k_i}(T^{-k_i}B_i)$ is
approximately $p^c_{N-k_i}(A)\cdot\nu_0(T^{-k_i}B_i)/
\nu_0(A)$ due to (25), (26). This is an approximation with
an exponentially small error, precisely
$$
  \left|\frac{p^c_{N-k_i}(T^{-k_i}B_i)\nu_0(A)}
        {p^c_{N-k_i}(A)\nu_0(T^{-k_i}B_i)}
          -1 \right| \leq c_{25} \alpha_9^m,           \tag 27
$$
where $m$ is the order of homogeneity of the 
parallelogram $A$, as in (26). We now obtain
$$
   p^c_N(\cup B_i)=\sum_i p^c_{N-k_i}(T^{-k_i}B_i)
   = (1+\Delta_m)(\nu_0(A))^{-1}\sum_i p^c_{N-k_i}(A)
           \nu_0(T^{-k_i}B_i).
$$
The error term $\Delta_m$ here is exponentially small in $m$,
as stated in (27). 

   Next we estimate the value $p^c_N(B\setminus
\cup B_i)$. First, the $p^c_N$-measure of the set 
$(B\setminus(B^{(0)}\cup(\cup B_i)))$ is exponentially small
in $n_1-n_0$ due to Lemmas 10-15, because that set consists
of the points $y\in B$ such that $T^{-n_1}y,\ldots,T^{-n_0}y$ 
do not belong to $A$. The estimation of the value $p^c_N(B^{(0)})$ 
is based on the following lemma.

\proclaim{Lemma 16 (Bound for irregular parts)}
$p^c_N(\frak I(T^{-n}B\cap A))\leq c_{26}\alpha_{10}^n$
for certain $\alpha_{10}\in (0,1)$ and all $N\geq 1, n\geq 1$.
\endproclaim

   The proof of Lemma 16 is essentially the same as
that of Proposition 5.2 in [9]. The only difference
is that we use here the measure $p^c_N$ instead of
the invariant smooth measure on $M$. The validity
of this change of measures is justified by our Remark 9.

Summarizing the above estimates, we obtain the decomposition
$$
   p^c_N(B)= (1+\Delta_m)(\nu_0(A))^{-1}\sum_i p^c_{N-k_i}(A)
           \nu_0(T^{-k_i}B_i)+\Delta_0+\Delta_1,
                      \tag 28
$$
where $|\Delta_1|<c_{27}\alpha_{11}^{n_1-n_0}$ and
$|\Delta_0|<c_{28}\alpha_{12}^{n_0}$ for certain 
$\alpha_{11},\alpha_{12}\in (0,1)$. 
 
   The remarkable formula (28) allows us to estimate the
measure $p^c_N(B)$ for an arbitrary homogeneous maximal 
parallelogram $B$. Denote $\varepsilon=\nu_0(B)$, then 
$\rho(\gamma^s_B(y)) \geq c_{29}\varepsilon$ for any $y\in B$.
Now we choose $n_0=-C_0\ln\varepsilon$ and $n_1=-C_1\ln
\varepsilon$ with some large constants $C_0<C_1$. Then 
both $|\Delta_0|$ and $|\Delta_1|$ in (28) do not exceed 
$\varepsilon^D$ with some large $D>0$. On the other hand, 
if the difference $C_1-C_0$ is also large enough, then the 
majority of points of $B$ are transformed into $A$ under 
the map $T_{\ast}^{(-)(n_1-n_0)}\circ T_E^{-n_0}$, so that 
$\nu_0(\cup B_i)$ will certainly be close to $\varepsilon$.
To estimate the values $\nu_0(T^{-k_i}B_i)$ we observe
that 
$$
    \ln (\nu_0(T^{-k_i}B_i)/\nu_0(B_i))\leq\omega En_1=
    -C_1\omega E\ln\varepsilon
$$
due to (19). This implies the estimate
$$
 \varepsilon^{1+C_1\omega E}\leq\sum_i\nu_0(T^{-k_i}B_i)
           \leq\varepsilon^{1-C_1\omega E}.
                    \tag 29
$$
We obtain for small $E$ that the first term in the RHS
of (28) is actually the principal one and we can neglect
the others. It is also useful to note how the singularity
of the limit measure $\nu^+_E$ can arise. The inequalities
(29) imply
$$
  (\nu_0(B))^{C_1\omega E}\leq\frac{p^c_N(B)}{\nu_0(B)}
        \leq(\nu_0(B))^{-C_1\omega E}.
                      \tag 30
$$
Thus the ``density" of $p^c_N$ with respect to $\nu_0$
can approach either zero or infinity as $N\to\infty$ depending
on which of two processes overcomes: the contracting or
the expanding. Due to (19) the contracting prevails when
the particle with the initial conditions in $B$ travels
mainly in the direction of the field $\bE$. The expanding
prevails when the particle travels in the opposite direction.
The displacement of the particle in the perpendicular direction
causes no effect on the density of $p^c_N$.

Remark. In our model the particle has enough freedom to travel
along or opposite to the field direction. If the billiard table 
is closed or extended only in the perpendicular direction to
the applied field, then the particle has no such freedom 
and the density of $p^c_N$ stays uniformly bounded. Although
that density apparently oscillates as $N$ grows, the limit measure
$\nu_E^+$ nevertheless exists and is absolutely continuous
with respect to the Lebesgue measure. Our response theory
(Propositions 1 - 4) is formally correct but trivial since
the main constants $\bD,\bsigma$ and the current $\bJ$ are 
all zeroes.
 
   As an immediate consequence of the decomposition (28)
and the above remarks we obtain that $\nu_E^L(B)\geq
C(B)>0$ as soon as $\nu_0(B)>0$. Here $C(B)$ is independent
of the initial fiber $\gamma^u$ and of 
the value of $E$, provided the latter is small enough.
Moreover, the decomposition (28) implies
$$
    \frac{\nu_E^U(B)}{\nu_E^L(B)}\leq
         (1+\Delta^{\prime}_m)\frac{\nu_E^U(A)}{\nu_E^L(A)},
             \tag 31
$$
where the constant $\Delta^{\prime}_m$ is determined
by $A$ alone and approaches zero as $m\to\infty$.

   So far we have applied only ``local" arguments studying
the evolution of a particular parallelogram $B$. These have
given us only a ``rough" estimate (31). Next we are
going to show that actually $\nu_E^U(B)=\nu_E^L(B)$
and thus this value determines $\nu_E^+(B)$. To
this end we have to involve certain ``global"
arguments. Namely, we use the Markov sieve
$\frak S_n$ for some large $n$ and study the
joint evolution of all its parallelograms. It
can be well approximated by a probabilistic
Markov chain as is explained below.

   The properties of the Markov sieve $\frak S_n$
and our Lemma 11 yield the bound
$$
    p_N^c(\frak N(\xi_n))\leq c_{30}\alpha_{13}^n
           \tag 32
$$
for every large $N$, say, for $N\geq -C\ln\rho(\gamma^u)$
for some large $C>0$. Furthermore, we can easily estimate 
the $p_N^c$-measure of the set of points in the quadrilaterals 
$\Pi_1,\ldots,\Pi_I$ which do not belong to the parallelograms 
$A_1,\ldots,A_I$. These points have too short unstable
or stable fibers, so that Lemma 7, along with
the above estimate (32), gives the bound
$$
    p_N^c(M\setminus\cup A_i)\leq c_{31}\alpha_{14}^n.
           \tag 33
$$
In other words, the measure $p_N^c$ is almost
concentrated on the Markov sieve $\frak S_n$, up
to an exponentially small error term.

   Now denote $\pi_i(N)=p^c_N(A_i)$ and $\pi_{ij}^{(K)}
(N)=p^c_{N+K}(A_j\cap T^KA_i)/p^c_N(A_i)$. Setting
$A_0=M\setminus\cup A_i$ and letting the indices
$i,j$ in the above notations run from $0$ to $I$ we
make $\|\pi_i(N)\|$ a probability distribution
and $\|\pi_{ij}^{(K)}(N)\|$ a stochastic matrix.
The measure $p^c_N$ inside the quadrilateral $\Pi_i$ 
is concentrated on a finite union of unstable fibers 
$u$-inscribed in $\Pi_i$ which are images of $\gamma^u$
under $T_E^n$. Let $\tilde{\gamma}^u$ be one of those
fibers and $\tilde{\gamma}_i^u=\tilde{\gamma}^u
\cap A_i$. Denote $\tilde p^c$ the probabilistic
conditional measure induced by $\nu_0$ on 
$\tilde{\gamma}_i^u$ and $\tilde p^c_n=
T^n\tilde p^c$ for $n\geq 1$. To the set $\tilde
{\gamma}_i^u$ we can apply the above arguments
involving a fixed parallelogram $A$ and resulting
in the estimate (28). These arguments show again
that
$$
  \tilde p^c_K(B)=(1+\Delta_m)(\nu_0(A))^{-1}
  \sum_i\tilde p^c_{K-k_i}(A)\nu_0(T^{-k_i}B)
   +\Delta_0+\Delta_1,   \tag 34
$$
where $B$ stands for $A_j$. Choose, as in (28),
$n_0=C_0n, n_1=C_1n$ and $K=C_2n$ with sufficiently
large constants $C_0,C_1,C_2$ such that $C_1-C_0$ and $C_2-C_1$
are also large enough. Then again both $|\Delta_0|$ and
$|\Delta_1|$ in (34) do not exceed $\alpha_5^n$
for some small $c_{32}\alpha_{15}$ determined by $C_0$
and $C_1$.

   Comparing (34) to (28) we conclude that
$\tilde p^c_K(B)\geq \frac12 p^c_N(B)$ provided
$C_0,C_1,C_2$ and $N$ are large enough. As
a result we obtain that
$$
   \pi_{ij}^{(K)}\geq\frac12\pi_j(N).
                 \tag 35
$$
Next, due to the $n$-homogeneity of the parallelogram
$A_i$ the values $\tilde p^c_K(B)$ are almost the
same for different fibers $\tilde{\gamma}^u$
$u$-inscribed in $\Pi_i$. To be specific, if
$\tilde{\tilde{\gamma}}^u$ is another fiber
of that kind, then
$$
  \left|\tilde{\tilde p}^c_K(B)/\tilde p^c_K(B)-1\right |
  \leq c_{33}\alpha_{16}^n.   \tag 36
$$
As a result, the values $\pi_{ij}^{(K)}(N)$ are almost
independent of $N$, and so we can find an approximative
stochastic matrix $\pi_{ij}^{(K)}$ such that 
$$
  \left|\pi_{ij}^{(K)}(N)/\pi_{ij}^{(K)}-1\right |
  \leq c_{34}\alpha_{17}^n   \tag 37
$$
for all $i,j\geq 1$. The estimate (36) yields also an
important Markovian property
$$
  \frac{p^c_{N+K}(A_j\cap T^KA_{i_1}\cap\cdots\cap T^{LK}A_{i_L})}
   {p^c_N(A_{i_1}\cap\cdots\cap T^{(L-1)K}A_{i_L})}=
   \pi_{i_1j}^{(K)}(N)(1+\Delta^{\prime}),
                         \tag 38
$$
where $|\Delta^{\prime}|\leq c_{35}\alpha_{18}^n$. Moreover,
$\pi_{i_1j}^{(K)}(N)$ in (38) can be replaced by
$\pi_{i_1j}^{(K)}$ due to the approximation (37).

   As a result we obtain an approximation of the
joint evolution of the parallelograms of the Markov
sieve $\frak S_n$ by a stationary Markov chain. To 
be specific,
$$
  \pi_j(N+LK)=(1+\Delta)\sum_{i_1,\ldots,i_L}
  \pi_{i_L}(N)\pi_{i_Li_{L-1}}^{(K)}\cdots\pi_{i_1j}^{(K)}
                     \tag 39
$$
with some $|\Delta|<c_{36}\alpha_{19}^n$, provided $L$ is
not too large, say, $L=n$. Now we have an approximative
stationary Markov chain (39) with the estimate (33)
for the total measure of the ``marginal" set $M
\setminus\cup A_i$ and with the regularity condition
(35) of Ibragimov type, see [22] and also [6]. 
These basic properties allow
us to estimate the rate of mixing in the Markov chain
and to prove a rapid convergence in $L$ of the
probability distribution $\|\pi_j(N+LK)\|$ to the
stationary distribution $\|\pi_j\|$ of the matrix
$\|\pi_{ij}^{(K)}\|$. The corresponding reasonings involve
typical estimates from the theory of Markov chains.
The proof is essentially the same as that of the
theorem 4.1 in [6], and we do not reproduce it here.
The actual results are
$$
     \sum_j |\pi_j(N+LK)-\pi_j|\leq c_{37}\alpha_{20}^n
                \tag 40
$$
and
$$
    \sum_j \left|\frac{p^c_{N+LK}(T^{LK}A_i\cap A_j)}{\pi_i}-\pi_j\right|
            \leq c_{38}\alpha_{21}^n.
                       \tag 41
$$
for a typical parallelogram $A_i$. This last statement
means that there is a subset $R_{\ast}\in\frak S_n$
such that (41) holds for {\it every} $A_i\in R_{\ast}$,
and the total $p^c_N$-measure of all the other
parallelograms, i.e. those in $\frak S_n\setminus R_{\ast}$,
is less than $c_{39}\alpha_{22}^n$.

   We are now able to prove the existence of the limit
measure $\nu_E^+$. First, for any quadrilateral $\Pi$
we prove that $\nu^U_E(\Pi)=\nu^L_E(\Pi)$. For large
$n\geq 1$ consider the Markov sieve $\frak S_n$
with elements $A_1,\ldots,A_2$. The measure
$\nu_E^U(\Pi\setminus\cup A_i)$ is small enough
due to (33). The parallelograms $A_i$ crossing
the boundary $\partial \Pi$ also have a small total
measure due to Remark 12. Therefore, the measures
$\nu_E^U$ and $\nu_E^L$ are concentrated mainly
on the union of parallelograms inside $\Pi$.
We denote this union by $\Pi_n$. By virtue of (40) the measure
$p^c_N(\Pi_n)$ is sufficiently close to the sum
of the values $\pi_i$ for the parallelograms included into $\Pi_n$.
Since this last sum is independent of $N$, we
obtain that at least $\nu_E^U(\Pi)/\nu_E^L(\Pi)\leq c_{40}
\alpha_{23}^n$. Finally, $n$ here can be choosen
arbitrarily large, so that actually $\nu_E^U(\Pi)=
\nu_E^L(\Pi)$.

   These arguments can be also extended to 
maximal parallelograms. A maximal parallelogram
$A$ can be obtained by removing from its
support $\Pi(A)$ an infinite number of smaller
quadrilaterals (gaps) ${\Pi_i}$, see e.g. [5,6,9] for detail.
The $\nu_0$-measures of those gaps decay exponentially
fast, see e.g. [9, Lemma B.1]. Combining this fact
with the estimate (30) we obtain the necessary
tail bound for the $\nu_E^U$-measures of those gaps,
and then prove the formula $\nu_E^+(A)=
\nu_E^+(\Pi(A))-\sum_i\nu_E^+(\Pi_i)$. Thus we establish
the existence of the measure $\nu_E^+$.

\noindent{\it (e) Decay of Correlations}

   The estimate (41) has not yet been used. It readily
yields the ergodicity of the measure $\nu_E^+$.
Moreover, we could as well establish a subexponential
rate of the decay of correlations with respect to that
measure, as for billiards in [6]. However, we do not 
need this exactly. What we really need in Section (a) 
is an estimate for the decay of correlations with 
respect to the measure $\nu_0$.

Let us fix here our definition of the H\"{o}lder classes $H_{\alpha}^*$,
for small $E$, by specifying the set of allowed discontinuity to be the 
singularity sets of the maps $T_E^{\pm}$ and $T_0^{\pm}.$ Then, 
for instance, the functions  $\tau_0(x)$ and ${ \bDelta}_0(x),$
as well as   $\tau_E(x)$ and ${ \bDelta}_E(x),$ 
belong to $H_{\alpha}^*$ (see below.)

\proclaim{Theorem 17 (Decay of correlations)}
 For any two functions $f,g\in H^{\ast}_{\alpha}$ 
such that
$$
     \nu_0(f)=0
$$
and for any $n\geq 1$ we have
$$
   \left|\nu_0(f\cdot(g\circ T_E^n))\right|\leq C(f,g)\lambda_1^{\sqrt n},
$$
where $\lambda_1\in (0,1)$ is a constant determined by $T$ and
$\alpha$.
\endproclaim

   The proof of Theorem 17 is based on the estimate (41)
along with supplementary estimates (33). It goes the same
way as the proof of the theorem 1.1 in [6] and we omit the
details.

   We need also certain estimates for the constants $\lambda_1$ 
and $C(f,g)$ in Theorem 17. 
These estimates readily come from the proof of Theorem 17
and were first explicitly given in [9]. The constant
$\lambda_1$ can be chosen as $\lambda_2^{\alpha}$ for
some $\lambda_2\in (0,1)$ which is independent of $\alpha$ 
and $E$. Furthermore, we can set $C(f,g)=(C_f+M_f)
(C_g+M_g)$, where $C_f$ is the factor in the
H\"older condition and $M_f=\max_M|f(x)|$.

   We now return to the specific function
$$
   f_E(x)=1-\exp[-\beta{ \bE\cdot\bDelta}_E(x)]
$$
which appears in Part I and section (a). This function lies
in a H\"{o}lder class  $H_{\alpha}^*$ and, in fact, 
the corresponding coefficients $C_f,M_f$
vanish at least linearly in $E$ as $E\to 0$.
To be specific,
$$
     |f(x)|\leq 2E\beta\Delta_{\max} \tag 42
$$
where $\Delta_{\max}=\max_E\max_{x\in M}|{ \bDelta}_E(x)|$ 
is finite due to the finiteness
of the horizon. Furthermore, if $x,y$ belong to the same
component of smoothness of $T_E$, then
$$
  |f_E(x)-f_E(y)|\leq 2E\beta|{ \bDelta}_E(x)-{ \bDelta}_E(y)|
  \leq 2E\beta C|x-y|^{1/2}, \tag 43
$$
where $C>0$ is independent of $E$. This last estimate
is easy to check for $E=0$, and then we apply the
$C^2$ closeness of $T_E$ to $T_0$ for small $E$. The estimate
(43) also gives the exponent $\alpha=1/2$ in the H\"older
condition. 

   The estimates (42) and (43) give a uniform
in $E$ bound for the decay of correlations which we
need in Section (a). Let $g_E$ be another function in $H^{\ast}_{\alpha}$
also depending on $E$ such that $C_g$ and $M_g$ are uniformly
bounded in $E$. Then
$$
   \left|\nu_0(f_E\cdot(g_E\circ T_E^n))\right|
                    \leq EC_0(f,g)\lambda_2^{\sqrt n}
$$
where $\lambda_2\in (0,1)$ does not depend on $E$ and $C_0(f,g)$
is uniformly bounded in $E$.

{\it (f) Entropy and fractal dimension}

   In this last section we prove the parts of Proposition 1 concerning
entropy and fractal dimension for our measure $\nu_E^+.$

   First, the Pesin formula expresses the measure-theoretic entropy
of the map $T_E$ as
$$
       h_{\nu_E^+} (T_E) = \tilde{\lambda}_E^u,
$$
where $\tilde{\lambda}_E^u$ is the positive Lyapunov 
exponent for $T_E$, which is $\nu_E^+$-a.e. constant 
in $M$ due to the ergodicity. This formula has
been proved in [24] for hyperbolic maps with singularities with the
only assumption that the underlying invariant measure is absolutely
continuous on unstable fibers, which is true in our case. The entropy
of the flow $S_E^t$ in the full space $\frak M$ is related to that of
the map $T_E$ through the well-known Abramov formula [1]:
$$
        h_{\mu_E^+} (\{S_E^1\}) = \bar{\tau}_E^{-1} h_{\nu_E^+} (T_E)
$$
Note that another expression for the entropy of a hyperbolic map
exists, which is in our notations 
$$
              h_{\nu_E^+} (T_E) = \nu_E^+(\tilde{\Lambda}_E^u)
                         \tag 44
$$
where $\tilde{\Lambda}_E^u(x)$ is the local exponential rate of
expansion for $x\in M$ under the map $T_E$; see also (18).

   As $E\to 0$, the function $\tilde{\Lambda}_E^u(x)$ converges
to $\tilde{\Lambda}_0^u(x)$ for every $x\in M\setminus(\cup S_n)$. 
Although that convergence is nonuniform, all those functions 
are uniformly bounded and continuous on their domains of 
definition. Furthermore, the measure $\nu_E^+$ weakly converges 
to $\nu_0$ as $E\to 0$ due to our estimate (30). Hence the RHS 
of (44) converges to $\nu_0(\tilde{\Lambda}_0^u)$ as $E\to 0$, 
so that
$$
        \lim_{E\to 0} h_{\nu_E^+} (T_E) = h_{\nu_0} (T_0).
$$
This is what we needed for the Corollary to Propositions 1 and 4
in Part I.

  Our estimates of the fractal dimension of the measure $\nu_E^+$
are based on the approach by L.-S.~Young. In her paper
[43] a chain of relations between the Hausdorff dimension $HD(\nu)$,
the capacity $C(\nu)$, the Renyi dimension $RD(\nu)$ and the entropy
$h(\nu)$ has been proven for an ergodic measure $\nu$ of a $C^{1+\alpha}$
diffeomorphism $T$ of a compact two-dimensional surface:
$$
     HD(\nu) = C(\nu) = RD(\nu)
          = h_{\nu}(T)\left[\frac1{\tilde{\lambda}^u}-\frac1{\tilde{\lambda}^s}\right].
                       \tag 45
$$  
Note that $\tilde{\lambda}^s$ is negative, so that the expression in
the brackets is always positive.
Our map $T_E$ is discontinuous, but its singularities are mild enough
to extend the proofs of (45) to $T_E$ along the lines of Katok-Strelcyn
[24]. The modifications are minor and we do not go into detail.

{\bf Acknowledgements.} N.C. and Ya. S. would like to thank
L.-S. Young for the suggestion that her dimension formula may be 
applicable to our problem.  G.E. and J.L.L. would like to thank
W. Hoover, D. J. Evans, and E. G. D. Cohen for introducing 
us to the Gaussian method and its applications in non-equilibrium 
molecular dynamics. All the authors thank G. Gallavotti for 
his many very helpful discussions concerning various aspects
of the whole subject. Finally, G. E. and J. L. L. wish to acknowledge
the support of the Air Force Office of Scientific Research, Grant No. 
AF-0010E-91, for its support while completing this work.

\enddocument